\documentstyle[11pt,aaspp4]{article}

\slugcomment{To appear in {\it Astrophys. J. Supp. Ser.}}

\begin{document}

\title{Dielectronic Recombination in Photoionized Gas.  II.  Laboratory
Measurements for Fe XVIII and Fe XIX}

\author{D. W. Savin and S. M. Kahn}
\affil{Department of Physics and Columbia
Astrophysics Laboratory, Columbia University, \\ New York, NY 10027, USA}
\authoremail{savin@astro.columbia.edu}
\author{J. Linkemann, A. A. Saghiri, M. Schmitt, M. Grieser, R. Repnow, 
D. Schwalm, and A. Wolf}
\affil{Max-Planck-Institut f\"{u}r Kernphysik, D-69117 Heidelberg, Germany\\
and Physikalisches Institut der Universit\"at Heidelberg, 69120
Heidelberg, Germany}
\author{T. Bartsch, C. Brandau, A. Hoffknecht, A. M\"{u}ller, and S. Schippers}
\affil{Institut f\"{u}r Kernphysik, Strahlenzentrum der
Justus-Liebig-Universit\"{a}t, D-35392 Giessen, \\ Germany}
\author{M. H. Chen}
\affil{Lawrence Livermore National Laboratory, Livermore, CA 94550, USA}
\and
\author{N. R. Badnell}
\affil{Department of Physics and Applied Physics,
University of Strathclyde, Glasgow, G4 0NG, United Kingdom}

\begin{abstract}

In photoionized gases with cosmic abundances, dielectronic
recombination (DR) proceeds primarily via $nlj \rightarrow
nl^\prime{j^\prime}$ core excitations ($\Delta n=0$ DR).  We have
measured the resonance strengths and energies for Fe XVIII to Fe XVII
and Fe XIX to Fe XVIII  $\Delta n=0$ DR.  Using our measurements, we
have calculated the Fe XVIII and Fe XIX $\Delta n=0$ DR rate
coefficients.  Significant discrepancies exist between our inferred
rates and those of published calculations.  These calculations
overestimate the DR rates by factors of $\sim 2$ or underestimate it by
factors of $\sim 2$ to orders of magnitude, but none are in good
agreement with our results.  Almost all published DR rates for modeling
cosmic plasmas are computed using the same theoretical techniques as
the above-mentioned calculations.  Hence, our measurements call into
question all theoretical $\Delta n=0$ DR rates used for ionization
balance calculations of cosmic plasmas.  At temperatures where the Fe
XVIII and Fe XIX fractional abundances are predicted to peak in
photoionized gases of cosmic abundances, the theoretical rates
underestimate the Fe XVIII DR rate by a factor of $\sim 2$ and
overestimate the Fe XIX DR rate by a factor of $\sim 1.6$.  We have
carried out new multiconfiguration Dirac-Fock and multiconfiguration
Breit-Pauli calculations which agree with our measured resonance
strengths and rate coefficients to within typically better than
$\lesssim 30$\%.  We provide a fit to our inferred rate coefficients
for use in plasma modeling.  Using our DR measurements, we infer a
factor of $\sim 2$ error in the Fe XX through Fe XXIV $\Delta n=0$ DR
rates.  We investigate the effects of this estimated error for the
well-known thermal instability of photoionized gas.  We find that
errors in these rates cannot remove the instability, but they do
dramatically affect the range in parameter space over which it
forms.

\end{abstract}


\section{Introduction}
\label{sec:Introdution}

Photoionized gases form in planetary nebulae, H II regions, stellar
winds, cold novae shells, active galactic nuclei, X-ray binaries, and
cataclysmic variables.  In such gases the electron temperature $T_e$ at
which an ion forms (\cite{Kall96a}) is far below that where the ion
forms in coronal equilibrium (\cite{Arna85a}; \cite{Arna92a}).  As a
result, the dominant electron-ion recombination processes are radiative
recombination (RR) and low temperature dielectronic recombination via
$nlj \rightarrow nl^\prime{j^\prime}$ excitations of core electrons
($\Delta n=0$ DR).  Also, X-ray line emission is produced not by
electron impact excitation but by RR and DR (\cite{Lied90a};
\cite{Kall96a})

Recent {\it ASCA} observations of the low-mass X-ray pulsar 4U 1626-67
(\cite{Ange95a}) and the X-ray binary Cygnus X-3 (\cite{Lied96a}) have
spectroscopically confirmed the low $T_e$ of photoionized gas and
demonstrated some of the unique properties of such gas.  The
soon-to-be-launched satellites {\it AXAF}, {\it XMM}, and {\it Astro-E}
are expected to collect spectra which will reveal, in even greater
detail, the X-ray properties of photoionized gases.  Of particular
interest will be $n\ge3 \rightarrow n=2$ line emission of Fe XVII to Fe
XXIV (the iron $L$-shell ions) which dominates the $0.7-2.0$ keV (6-18
\AA) bandpass.

Iron $L$-shell ions form over a wide range of physical conditions and
are expected to provide many valuable plasma diagnostics
(\cite{Kahn95a}).  However, the accuracies of these diagnostics will be
limited by uncertainties in the relevant atomic data.  This will be an
issue especially for low temperature DR rate coefficients of iron $L$
ions.  These rates are theoretically and computationally challenging as
they require accurate energy levels for ions with partially-filled
shells and involve calculating a near-infinite number of states.  The
challenge of these calculations can be seen by the spread in the
computed $\Delta n=0$ DR rates for Fe XVIII to Fe XVII and Fe XIX to
Fe XVIII.  Existing theoretical Fe XVIII (\cite{Rosz87a};
\cite{Chen88a}; \cite{Dasg90a}) and Fe XIX (\cite{Rosz87b};
\cite{Dasg94a}) rates differ by factors of 2 to 4 over the temperature
ranges where these ions are predicted to form in photoionized gas of
cosmic abundances (\cite{Kall96a}).

DR begins when a free electron excites an ion and is simultaneously
captured.  This state $d$ may autoionize.  DR is complete when $d$
emits a photon which reduces the energy of the recombined system to
below its ionization limit.  Conservation of energy requires
\begin{equation}
E_k=\Delta E - E_b, 
\label{eq:rescond}
\end{equation}
where $E_k$ is the kinetic energy of the incident electron, $\Delta E$
the excitation energy of the core electron in the presence of the
incident electron, and $E_b$ the binding energy released when the free
electron is captured.  Because $\Delta E$ and $E_b$ are quantized, DR
is a resonance process.

The strength of a DR resonance is given by the integral of the
resonance cross section over energy.  In the isolated resonance
approximation, the integrated strength of a particular DR resonance
$\hat{\sigma}_d$ can be approximated as (\cite{Kilg92a})
\begin{equation}
\label{eq:resonancestrength}
\hat{\sigma}_d={h{\cal R} \over E_d} \pi a_0^2 {g_d \over 2g_i}
{A_a(d\rightarrow i)\sum_f A_r(d\rightarrow f) \over
\sum_\kappa A_a(d\rightarrow \kappa) + \sum_{f^\prime} 
A_r(d\rightarrow f^\prime)}.
\end{equation}
Here $h$ is the Planck constant; ${\cal R}$ is the Rydberg energy
constant; $E_d$ is the energy of resonance $d$; $a_0$ is the Bohr
radius; $g_d$ and $g_i$ are the statistical weights of $d$ and of the
initial ion, respectively; $A_a$ and $A_r$ are the autoionization and
radiative decay rates, respectively; $\sum_f$ is over all states stable against
autoionization; $\sum_{f^\prime}$ is over all states energetically below $d$;
both $\sum_f$ and $\sum_{f^\prime}$ may include cascades through lower-lying
autoionizing states and ultimately to bound states; 
and $\sum_\kappa$ is over all states attainable by autoionization of
$d$.

To address the needs for modeling photoionized gases, we are carrying
out a series of experiments to measure the $\Delta n=0$ DR
rates for the iron $L$-shell ions.  Measurements are performed using
the heavy-ion Test Storage Ring (TSR) at the Max-Planck-Institute for
Nuclear Physics in Heidelberg, Germany (\cite{Habs89a};
\cite{Kilg92a}).  In Savin et al.  (1997), we gave a summary of our
measurements for Fe XVIII.  Here we present a more detailed analysis of
those results as well as our new measurements for Fe XIX.  Measurements
have also been carried out for $M$-shell Fe XVI (Linkemann et
al.\ 1995).

Fe XVIII is fluorinelike with a $2p_{3/2}$ hole and a ground state of
$^2P_{3/2}$.  Table~\ref{tab:energylevels} lists the energies (relative
to the ground state) of all Fe XVIII levels in the $n=2$ shell.  Fe
XVIII can undergo $\Delta n=0$ DR via the capture channels
\begin{eqnarray}
{\rm Fe}^{17+}(2s^2 2p^5[^2P_{3/2}]) + e^-
\: &\rightarrow \left\{ \begin{array}{ll}
{\rm Fe}^{16+}(2s^2 2p^5[^2P_{1/2}]nl) & (n=18,\ldots,\infty)\\
{\rm Fe}^{16+}(2s   2p^6[^2S_{1/2}]nl) & (n=6,\ldots,\infty).
\end{array} \right.
\end{eqnarray}
The first channel involves the excitation of a $2p_{1/2}$ electron to
the $2p_{3/2}$ subshell.  This fills the $2p_{3/2}$ subshell, creates a
hole in the $2p_{1/2}$ subshell, and leaves the ion core in a
$^2P_{1/2}$ state.  The second channel involves the excitation of a
$2s_{1/2}$ electron to the $2p_{3/2}$ subshell.  This fills the
$2p_{3/2}$ subshell, creates a hole in the $2s_{1/2}$ subshell, and
leaves the ion core in a $^2S_{1/2}$ state.  The radiative
stabilization of these autoionizing states to bound configurations of
Fe XVII leads to DR resonances for collision energies between 0 and
$\sim132$~eV.  The lowest energy $\Delta n=1$ DR resonances occur for
$E_k \sim 220$ eV.

Table~\ref{tab:energylevels} also lists the energies (relative to the ground
state) of all Fe XIX levels in the $n=2$ shell.  Fe XIX is oxygenlike
and can undergo $\Delta n=0$ DR via a number of channels.  Those
channels which are strong enough for us to observe DR resonances are
\begin{eqnarray} 
{\rm Fe}^{18+}(2s^2 2p^4[^3P_2]) + e^-
\rightarrow \left\{ \begin{array}{ll} 
{\rm Fe}^{17+}(2s^2 2p^4[^3P_0]nl) & (n=22,\ldots,\infty)\\ 
{\rm Fe}^{17+}(2s^2 2p^4[^3P_1]nl) & (n=20,\ldots,\infty)\\
{\rm Fe}^{17+}(2s^2 2p^4[^1D_2]nl) & (n=15,\ldots,\infty)\\ 
{\rm Fe}^{17+}(2s^2 2p^4[^1S_0]nl) & (n=11,\ldots,\infty)\\ 
{\rm Fe}^{17+}(2s 2p^5[^3P_2^o]nl) & (n=7,\ldots,\infty)\\  
{\rm Fe}^{17+}(2s 2p^5[^3P_1^o]nl) & (n=7,\ldots,\infty)\\  
{\rm Fe}^{17+}(2s 2p^5[^3P_0^o]nl) & (n=6,\ldots,\infty).
\end{array} \right. 
\end{eqnarray} 
Radiative stabilization of the Fe XVIII autoionizing states to bound
configurations of Fe XVIII leads to measurable DR resonances for
electron-ion collision energies between 0 and $\sim 128$~eV.  The
lowest energy $\Delta n=1$ DR resonances occur for $E_k \sim 218$ eV.

In Section \ref{sec:ExperimentalTechnique} we describe the experimental
arrangement used to obtain the present results.  Section
\ref{sec:ResultsandDiscussion} compares our measurements with published
DR calculations.  In Section \ref{sec:Theory} we discuss new
theoretical calculations which we have carried out for comparison with
our measurements, while  Section \ref{sec:AstrophysicalImplications}
discusses the astrophysical implications of our results.

\section{Experimental Technique} 
\label{sec:ExperimentalTechnique}

DR measurements are carried out by merging, in one of the straight
sections of TSR, the circulating ion beam with an electron beam.  After
demerging, recombined ions are separated from the stored ions using
a dipole magnet and directed onto a detector.  The relative
electron-ion collision energy can be precisely controlled and the
recombination signal measured as a function of this energy.  A detailed
description of TSR (\cite{Habs89a}) and the procedures used for DR
measurements have been given elsewhere (\cite{Kilg92a};
\cite{Lamp96a}).  The experimental arrangement for our Fe XVIII
measurements is discussed in Savin et al.\ (1997).  Here we describe
primarily the setup used for our Fe XIX results and mention only those
details for Fe XVIII which were not discussed previously.

Negative $^{56}$Fe ions are accelerated and stripped using a tandem
accelerator and then further accelerated to 241 MeV and stripped to
their final charge state of 18+.  The ions are injected into TSR and
accumulated using repeated multiturn-injection stacking techniques
(\cite{Grie91a}) and electron cooling (\cite{Poth90a}).  In this manner
stored ion currents of $\sim 20-60$ $\mu$A are achieved.  The storage
lifetime is $\sim 30$ s.  After stacking, the ions are electron cooled
for $\sim 5$ s before data collection begins.  This is long compared to
the lifetime of the various metastable Fe XIX levels (\cite{Chen79a}),
and the ions are assumed to be in their ground state when the DR
measurements begin.  The beam width, measured using a beam profile
monitor (\cite{Hoch94a}), is $\sim 2-3$ mm after cooling.

The electrons are guided by a magnetic field of 41 mT and merged with
the ions over a straight interaction region of length $L \sim 1.5$ m.
For cooling, the electron velocity, $v_e$, is matched to that of the
ions, $v_i$.  The electron beam is adiabatically expanded before
merging (from a diameter of $\sim 0.95$ to $\sim 3.4$ cm) to reduce its
velocity spread perpendicular to the magnetic field (\cite{Past96a}).
The resulting energy distribution of the electrons is best described in
the present experiment by an anisotropic Maxwellian distribution
characterized by temperatures of $k_BT_\bot \sim 17$ meV and $k_BT_\|
\sim 0.4$ meV which are, respectively, perpendicular and parallel to
the confining magnetic field (and $k_B$ is the Boltzmann constant).
The electron density $n_e$ varies between $\sim 2.7$ and $5.1 \times
10^7$ cm$^{-3}$.

For the Fe XIX measurements, the electron energy is chopped from
cooling to a reference energy and then to the measurement energy.  Each
energy step is maintained for 25 ms.  After waiting for 5 ms, data are
acquired for the last 20 ms of each step.  The reference energy is
chosen so that RR and DR contribute insignificantly to the
recombination signal.  The recombination signal at the reference energy
represents only the background caused by charge transfer (CT) of the
ions with the rest gas in TSR.  For the present results, the reference
energy is $\sim 1600$ eV greater than the cooling energy ($\sim 2360$
eV).  This corresponds to a center-of-mass energy of $\sim 170$ eV.
For the Fe XVIII measurements the same timing was used but there was no
step to a reference energy, i.e. the electron energy was chopped only
between cooling and measurement energy.

The relative electron-ion collision energies $E$ are calculated using
$v_e$ (as determined using the calibrated
electron acceleration voltage and accounting for space charge
effects in the electron beam) and $v_i$ in the
overlap region.  The resulting experimental energy scale was verified
by comparing the measured DR resonance energies to the calculated
resonance energies using
\begin{equation}
\label{eq:e_nl}
E_{nl} = \Delta E - \Biggl({z\over n-\mu_l}\Biggr)^2 {\cal R}
\end{equation}
where $z$ is the charge of the ion before recombination, $n$ is the
Rydberg level into which the free electron is captured, and $\mu_l$ is
the quantum defect for the recombined ion.  The quantum defect is to
account for energy shifts of those $l$ levels which have a significant
overlap with the ion core and cannot be described using the uncorrected
Rydberg formula.  For high enough $l$ levels this overlap is
insignificant.  Note that here the quantum defects are for recombined
ions with an excited core, not one in the ground state

To verify the Fe XVIII energy scale we used Equation \ref{eq:e_nl}
to fit the measured resonance energies for a given Rydberg series (see
Section \ref{sec:ResultsandDiscussion}).  Only resonances for which
$\mu_l$ is essentially 0 were used, and $\Delta E$ and ${\cal R}$ were
fit for.  For the Fe XVIII data we found relative differences between
the measured and calculated resonance energies on the order of
2\%.  These differences could be traced to small deviations of the
acceleration voltage from its calibrated values during the chopping
cycles.  In particular, on chopping from the lower-lying cooling energy
to the measurement energy, the electron
energy did not reach its desired value but remained below it by a small
amount.  To correct for these deviations we reduced the experimental
energy scale by a factor of $\sim 1.02$.  After this correction, a fit
of the measured resonance energies, using Equation \ref{eq:e_nl},
yielded a value of ${\cal R}$ which matched its known value and values
for $\Delta E$ which matched the spectroscopically measured energies of
the Fe XVIII $2s^22p^5(^2P_{3/2}) - 2s^22p^5(^2P_{1/2})$ and
$2s^22p^5(^2P_{3/2}) - 2s2p^6(^2S_{1/2})$ transitions (\cite{Suga85a};
\cite{Shir90a}).  We also compared the measured
energies for high-$n$, high-$l$ DR resonances and those predicted by
Equation \ref{eq:e_nl}.  The uncertainty in the corrected energy scale
is estimated to be $\lesssim 0.4$\%.

For the Fe XIX results, similar deviations of the acceleration voltage
from its calibrated values are found.  In particular, on chopping from
the higher-lying reference energy to the measurement energy, the
electron energy did not reach its desired value but stayed above it by
a small amount.  Technical reasons for the occurence of these voltage
errors in the Fe XVIII and the Fe XIX runs (in contrast to earlier DR
measurements at TSR) have been identified only during the course of the
data reduction after the measurements had been completed.  For the Fe
XIX run the discrepancy between measured and calculated resonance
energies was greatest for large energy differences between the
reference and measurement energies and was noticed because the energies
of the DR resonances for $v_i > v_e$ and for $v_i < v_e$ were not
symmetric around $v_i = v_e$.  The discrepancy became insignificant
near the $^3P_2 - ^3P^o_{1,2}$ DR series limits.  A small increase to
the electron energy assumed in the data analysis symmeterized the DR
resonance energies around $v_i = v_e$ but resulted in an overestimate
of the energy scale.

To re-calibrate our Fe XIX energy scale we used Equation \ref{eq:e_nl}
to fit the measured resonance energies for a given Rydberg series (see
Section \ref{sec:ResultsandDiscussion}).  Only those levels were used
for which $\mu_l$ is essentially 0, and $\Delta E$ and ${\cal R}$ were
fit for.  We then reduced the experimental energy scale by a factor of
$\sim 1.02$ so the Rydberg value matched its known value and the
inferred values for $\Delta E$ matched their spectroscopically measured
energies of the Fe XIX $2s^22p^4(^3P_2)-2s^22p^4(^3P_1)$,
$2s^22p^4(^3P_2)-2s^22p^4(^1D_2)$, $2s^22p^4(^3P_2)-2s2p^5(^3P_2^o)$,
and $2s^22p^4(^3P_2)-2s2p^5(^3P_1^o)$ transitions (\cite{Suga85a};
\cite{Shir90a}).  We also compared the measured
energies for high-$n$, high-$l$ DR resonances and those predicted by
Equation \ref{eq:e_nl}.  The uncertainty in the corrected energy scale
is estimated to be $\lesssim 0.7$\%.

Toroidal magnets are used to merge the ion and electron beams, and
after the straight interaction region to separate the beams again.  For
recombination measurements of Fe XVIII (Fe XIX), the motional electric
field produced by the second toroidal magnet can field ionize electrons
that after the DR process remain in Rydberg levels $n \gtrsim
n_{cut1}=145(132)$.  Correction dipole magnets after the electron
cooler can ionize electrons in Rydberg levels $n \gtrsim
n_{cut2}=143(130)$.  For the Fe XVIII data the magnetic field strengths
in the toroid and the correction dipoles were smaller than for the Fe
XIX data.  Downstream of the correction dipoles, recombined ions are
separated from the stored ions using another dipole magnet and directed
onto a fast scintillator, heavy-ion detector with an efficiency of
$\gtrsim 95$\% (\cite{Mier96a}).  For the Fe XVIII data the magnetic
field strength in this dipole magnet was larger than for the Fe XIX
data.  Electrons in Rydberg states with $n \gtrsim n_{cut3}= 56(63)$
can be field ionized by this magnet.  However, during the $\sim 5.1$ m
from the center of the cooler to the dipole magnet, electrons in high
Rydberg levels can radiatively decay below the various values of
$n_{cut}$.  Using the $\sim 156(177)$ ns flight time of the ions, the
fact that dielectronic capture occurs predominantly into $l \lesssim
8$, the hydrogenic formula for radiative lifetimes of Marxer \& Spruch
(1991), and the values of $n_{cut1}$ to $n_{cut3}$ for the Rydberg
cutoffs of the toroid and dipole correction magnets, we estimate that
DR into $n \lesssim n_{max} = 124(130)$ will radiatively decay below
the different values of $n_{cut}$.  The value of $n_{max}$ determines
the maximum quantum number of the DR-populated Rydberg level that can
be detected in our experimental arrangement.

The measured recombination signal is the sum of RR, DR, and CT off the
rest gas in the cooler.  Recombination of Fe XVIII due to CT is taken
into account by subtracting a constant count rate per ion such that the
measured rate coefficient at 134 eV matches the very low theoretical RR
rate at that energy (Lampert et al.\ 1996).  Since the pressure in the
cooler varies with the electron energy, there is a weak dependence of
the CT signal on the measurement energy.  Thus, the subtraction
technique used for the Fe XVIII data can remove most but not all of the
CT background signal.  The remaining CT signal, however, is a smooth
function of energy and can be readily subtracted out when extracting
resonance strengths from the data.

The Fe XIX recombination signal rate $R$ is calculated by subtracting
the rate at the reference energy from the rate at measurement.  Effects
of slow pressure variations during the scanning of the measurement
energy are therefore eliminated.  Only a weak contribution due to CT
remains in $R$ due to small fast pressure variations associated with
the chopping of the electron energy.  The measured recombination rate
coefficient $\alpha_L$ is given by $\alpha_L(E)=R\gamma^2/(L n_e
N_i/C)$ where $N_i$ is the number of ions stored in the ring, $C=55.4$
m the circumference of the ring, $\gamma^2=[1-(v_i/c)^2]^{-1} \approx
1.01$, and $c$ the speed of light.  The measured rate coefficient is a
convolution of the DR and RR cross sections with the experimental
energy spread, which is best described by an anisotropic Maxwellian
distribution in the co-moving frame of the electron beam (see above),
sitting atop the residual CT background.

Peaks in the measured data $\alpha_L(E)$ are due to DR.  As described
in Section \ref{sec:ResultsandDiscussion}, resonance strengths can be
extracted after subtracting a smooth background which is due to RR and
CT.  While this smooth contribution is dominated by RR at low collision
energies, we are unable to extract reliable RR rate coefficients from
the Fe XVIII and Fe XIX data due to remaining CT contributions in the
measured signal rate.

Systematic uncertainties for the absolute DR rate coefficients are due
to the ion current and electron density determinations, corrections for
merging and demerging of the electron and ion beams, recombined ion
detection efficiency, and uncertainties in the shape of the
residual CT background.  The total systematic uncertainty is estimated to
be less than 20\%.  Relative uncertainties for comparing DR rate
coefficients at different energies are estimated to be less than 10\%.
Uncertainties are quoted at a confidence level believed to be
equivalent to a 90\% counting statistics confidence level.

\section{Results and Discussion}
\label{sec:ResultsandDiscussion}

\subsection{Measured Resonance Strengths, Energies, and Quantum Defects}

\begin{figure}
\plotone{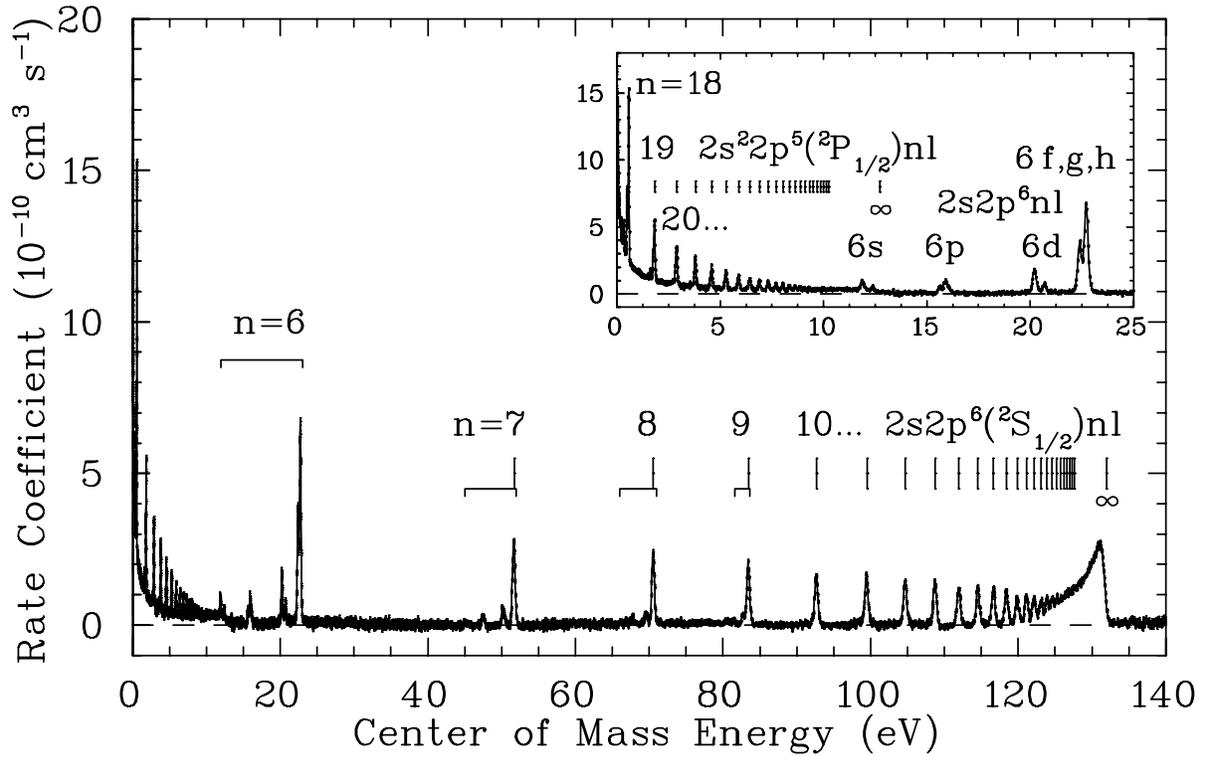}
\caption{Measured Fe XVIII to Fe XVII recombination rate coefficient
versus electron-ion collision energy.  $\Delta n=0$ DR resonances
resulting from $^2P_{3/2}-\ ^2P_{1/2}$ and $^2P_{3/2}-\ ^2S_{1/2}$ core
excitations are labeled.  The nonresonant ``background'' rate is due
primarily to RR with some residual CT (see text).}
\label{fig:FeXVIIIresonances}
\end{figure}

\begin{figure}
\plotone{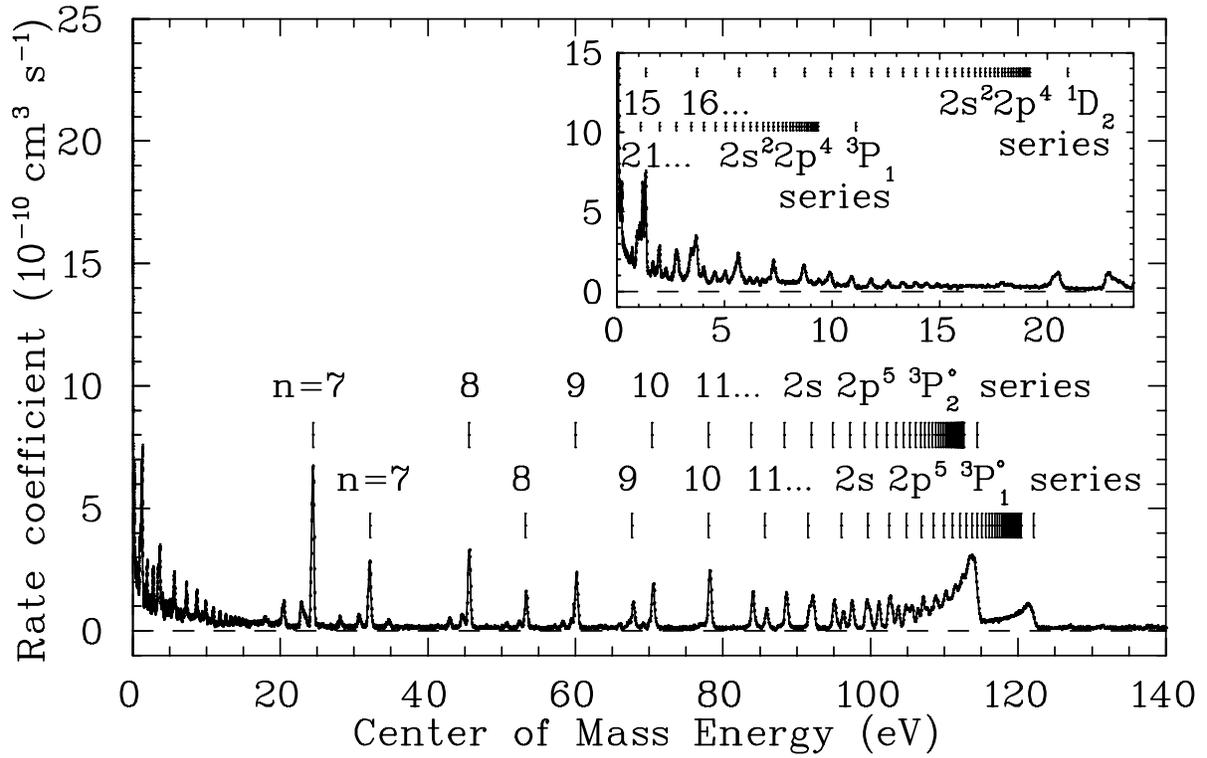}
\caption{Measured Fe XIX to Fe XVIII recombination rate coefficient
versus electron-ion collision energy.  $\Delta n=0$ DR resonances
resulting from $^3P_2-\ ^3P_1$, $^3P_2-\ ^1D_2$, $^3P_2-\ ^3P_2^o$, and
$^3P_2-\ ^3P_1^o$ core excitations are labeled for capture into high $l$
levels.  The nonresonant ``background'' rate is due primarily to RR
with some residual CT (see text).}
\label{fig:FeXIXresonances}
\end{figure}

Figures \ref{fig:FeXVIIIresonances} and \ref{fig:FeXIXresonances} show,
respectively, our measured Fe XVIII to Fe XVII  and Fe XIX to Fe XVIII
$\Delta n=0$ DR rate coefficients as a function of collision energy.
The resonances seen in these figures are due to the convolution of the
DR cross sections with the anisotropic Maxwellian electron
distributions of the two experiments.  Effects from the merging and
de-merging of the electron and ion beams have been corrected for as
described in Lampert et al.\ (1996).  In Figure
\ref{fig:FeXVIIIresonances}, at low energies DR of Fe XVIII via the
fine structure $^2P_{3/2}-\ ^2P_{1/2}$ core excitation can be seen.
Similar behavior was observed in an earlier measurement on the
isoelectronic Se XXVI (\cite{Lamp96a}).  For Fe XIX at low energies
(Figure \ref{fig:FeXIXresonances}), DR via the fine structure
$^3P_2-\ ^3P_1$ and $^3P_2-\ ^1D_2$ core excitations can clearly be
seen.  At high energies, resonances are visible due to
$^2P_{3/2}-\ ^2S_{1/2}$ excitations for Fe XVIII, and due to
$^3P_2-\ ^3P_2^o$ and $^3P_2-\ ^3P_1^o$ excitations for Fe XIX.

Resonance strengths and energies have been extracted by fitting the
measured resonances using the predicted asymmetric line shape
(\cite{Kilg92a}) for energies below 13 eV for Fe XVIII and below 25 eV
for Fe XIX.  Above 13(25) eV, the asymmetry is insignificant and we
have used Gaussian line shapes.  Tables \ref{tab:FeXVIIItable1} and
\ref{tab:FeXVIIItable2} list the extracted resonance energies and
strengths for Fe XVIII DR via the $^2P_{3/2} -\ ^2P_{1/2}$ and
$^2P_{3/2} -\ ^2S_{1/2}$ core excitations, respectively.  Measured Fe
XIX resonance energies and strengths are listed in Table
\ref{tab:FeXIXtable1}.  All energies quoted have been corrected as
described in Section \ref{sec:ExperimentalTechnique}.

Using Equation \ref{eq:e_nl} with the correct values of $\Delta E$ and
${\cal R}$ and the measured resonance energies in Tables
\ref{tab:FeXVIIItable1}, \ref{tab:FeXVIIItable2}, and
\ref{tab:FeXIXtable1}, we have determined the quantum defects for $s$,
$p$, and $d$ electrons of Fe XVII and for $p$ and $d$ electrons of Fe
XVIII (Table \ref{tab:qd}).  For Fe XVII, we do not use Fe XVIII DR
resonances via the $^2P_{3/2} -\ ^2S_{1/2}$ core excitation for capture
into the $n=6$ level.  For Fe XVIII, we use only those $n\ge8$ Fe XIX
DR resonances for the $^3P_2-\ ^3P_1^o$ and $^3P_2-\ ^3P_2^o$ core
excitations which are unblended.  We do not use resonances due to
capture into the $n\le6(7)$ level because the resonance structure is
too complicated to be accurately approximated using Equation
\ref{eq:e_nl}.  For each ion we have determined quantum defects
for two different excited core configurations.  For a
given value of $l$, one would expect different quantum defects for the
different cores.  However, due to the $\lesssim 0.4\%$ and
$\lesssim 0.7\%$ accuracy of our energy scale for Fe XVIII and Fe XIX
DR, respectively, we are unable to discern any difference.  For example,
the $\sim 10$\% difference between the measured values of $\mu_d$
for Fe XVII yields an $\lesssim 0.2$\% difference in
calculated resonance energies.

Theodosiou, Inokuti, \& Manson (1986) have calculated quantum defects
for ions with a ground state core (Table \ref{tab:qd}).  Their values
are consistently lower than the experimental values for Fe XVII and in
better agreement for Fe XVIII, but overall yield resonance energies
which agree with the measured values to within the uncertainty of our
energy scale.  This suggests that $\Delta n=0$ core excitations do not
significantly affect quantum defects for outer electrons in $n\ge7$
levels for Fe XVII and $n\ge8$ levels for Fe XVII.

\subsection{Inferred Maxwellian-Averaged Rate Coefficients and 
Comparison with Published Calculations}

\subsubsection{Fe XVIII}

\begin{figure}
\plotone{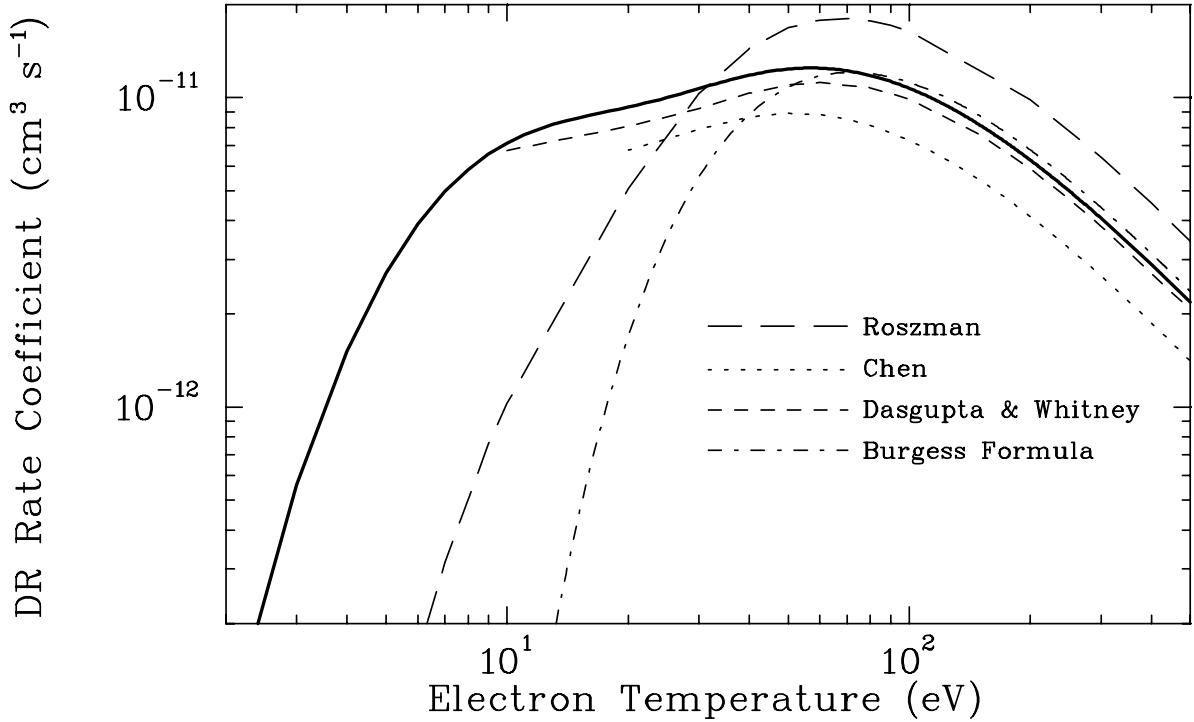}
\caption{Fe XVIII to Fe XVII Maxwellian-averaged rate coefficients for
$\Delta n=0$ DR via $^2P_{3/2}-\ ^2S_{1/2}$ core excitations.  The
thick solid line is the integration of the experimental DR resonance
strengths and energies extracted from the results shown in Figure
\protect\ref{fig:FeXVIIIresonances} and listed in Table
\protect\ref{tab:FeXVIIItable2}.  There is an estimated $\lesssim 20\%$
total systematic uncertainty in our experimentally inferred rate.
Calculations are from Roszman (1987a; long-dashed curve), Chen (1988;
dotted curve), Dasgupta \& Whitney (1990; short-dashed curve) and the
Burgess formula (1965; short-dashed-dot curve).}
\label{fig:FeXVIIIrates2}
\end{figure}

As shown in Savin et al.\ (1997), existing Fe XVIII $\Delta n=0$ DR
calculations do not account for DR via the $2p_{1/2} \rightarrow
2p_{3/2}$ (i.e., $^2P_{3/2} -\ ^2P_{1/2}$) fine-structure core
excitation.  For comparison with published theory, we have calculated a
Maxwellian-averaged Fe XVIII DR rate using only our measured $^2P_{3/2}
-\ ^2S_{1/2}$ DR resonance strengths and energies.  This rate is shown
in Figure \ref{fig:FeXVIIIrates2} together with the theoretical results
of Chen (1988), Roszman (1987), Dasgupta \& Whitney (1990), and
the Burgess (1965) formula using the oscillator strengths of Fuhr,
Martin, \& Wiese (1988).

Significant discrepancies exist between our inferred rate and the
calculations of Chen and Roszman.  The fully-relativistic,
multiconfiguration Dirac Fock (MCDF) calculations by Chen underestimate
the DR rate by a factor of $\sim 1.5$.  This may be partly due to
approximations which ignore DR for capture into Rydberg levels with
$l>8$ and partly due to the over-estimation of the resonance energies.
Including these levels and reducing the resonance energies would
increase the calculated DR rate.  Chen carried out explicit
calculations for $n \le 20$ and thus included the effects of
autoionization via a $2s2p^6(^2S_{1/2})nl \rightarrow
2s^22p^5(^2P_{1/2}) +e^-$ transition ($n \ge 18$).

The single-configuration, $LS$-coupling calculations by Roszman
overestimate the DR rate by a factor of $\sim 1.6$ for $k_BT_e \gtrsim
40$ eV.  This may be partly due to using $LS$-coupling which leaves out
the $2s2p^6(^2S_{1/2})nl \rightarrow 2s^22p^5(^2P_{1/2}) +e^-$
autoionization channel.  This opens up at $n=18$ and would, if
included, reduce the DR rate.  The discrepancy may also be partly due
to a possible error in the calculated resonance energies.  This could
also explain the low temperature behavior of Roszman's results.  Below
$\sim 40$ eV Roszman underestimates the $^2P_{3/2} -\ ^2S_{1/2}$ DR
rate because he calculated that DR via this channel becomes
energetically possible at $n=7$.  Our experiment shows this channel, in
fact, opens up for $n=6$.  Roszman also, like Chen, did not include
contributions due to capture into $l>8$ levels.  This results in
underestimating the DR rate.  In short, the full reason for the
discrepancy between Roszman's and our inferred rates is unclear.  The
differences between Chen's and Roszman's calculations may be partly
related to differences between MCDF and single-configuration,
$LS$-coupling methods.

The Burgess (1965) formula underestimates the DR rate for $k_BT_e
\lesssim 80$ eV. This is due to setting all DR resonance
energies to the threshold energy for the core excitation under
consideration.  This is valid only for DR into high $n$ levels; and
as noted by Burgess, himself, the formula is only applicable when
recombination in high $n$ levels dominates the DR process.

The agreement between our rate and the single-configuration,
intermediate-coupling calculations of Dasgupta \& Whitney is probably
serendipitous.  They carried out explicit calculations only for $n \le
15$ and $l \le 8$ and used extrapolation techniques for higher $n$
levels.  This leaves out the $2s2p^6(^2S_{1/2})nl \rightarrow
2s^22p^5(^2P_{1/2}) +e^-$ autoionization channel which results in an
overestimation of the DR rate.  Accounting for $l \le 8$ results in
an underestimate of the DR rate.  The agreement may be due to the various
approximations used roughly canceling out in the energy-averaged, total
rate coefficient.

This case clearly illustrates that comparisons of only
Maxwellian-averaged rate coefficients cannot be used to distinguish
between different theoretical techniques.  A detailed comparison
between experimental and theoretical resonance strengths and energies
is the only unambiguous way to verify the accuracy of DR rate
coefficient calculations.  This is now possible using high-resolution
DR measurements carried out at heavy-ion storage rings (as will be
illustrated in Section \ref{sec:Theory}).

\begin{figure}
\plotone{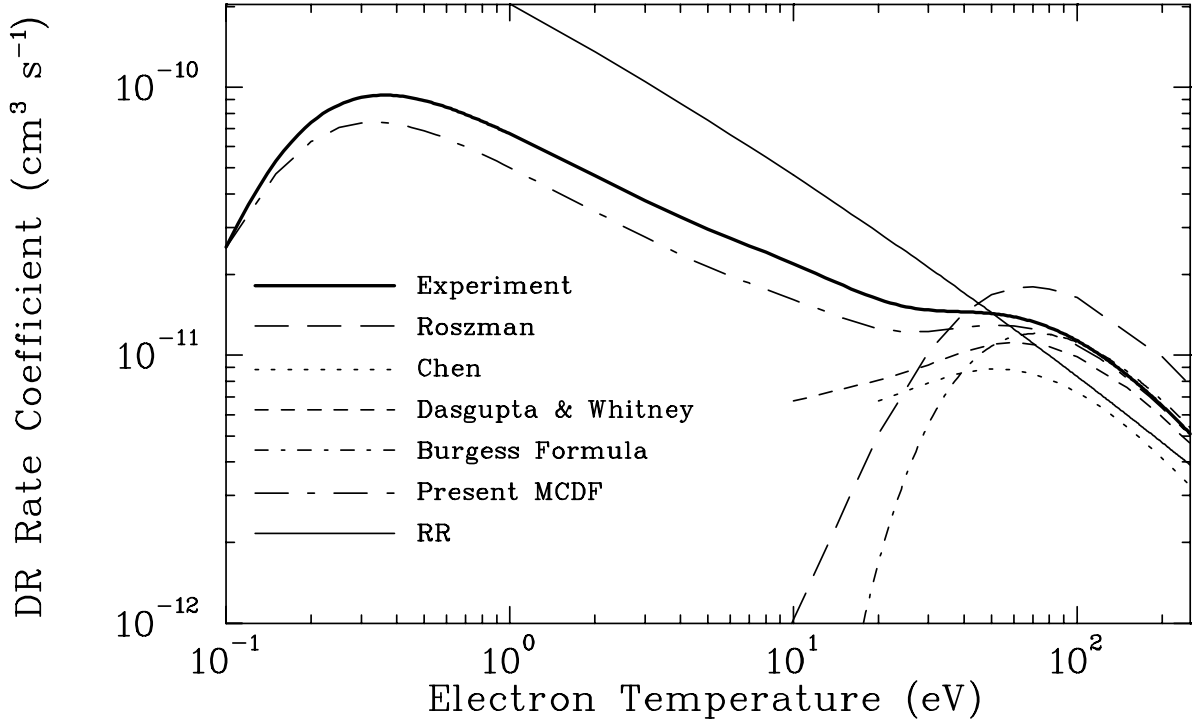}
\caption{Fe XVIII to Fe XVII Maxwellian-averaged $\Delta n=0$ DR total
rate coefficients.  The thick solid curve is the integration of the
experimental DR resonance strengths and energies extracted from the
results shown in Figure \protect\ref{fig:FeXVIIIresonances} and listed
in Tables \protect\ref{tab:FeXVIIItable1} and
\protect\ref{tab:FeXVIIItable2}.  There is an estimated $\lesssim 20\%$
total systematic uncertainty in our experimentally inferred rate.
Existing calculations by Roszman (1987a; long-dashed curve), Chen
(1988; dotted curve), Dasgupta \& Whitney (1990; short-dashed curve)
and the Burgess formula (1965; short-dashed-dot curve) do not include
the ${^2P}_{3/2}-{^2P}_{1/2}$ DR channel.  The long-dashed-dot curve
shows the results of our new MCDF calculations which include this
channel.  Our MCBP rate (not shown here) agrees well with our MCDF
rate.  The thin solid curve shows the recommended RR rate of Arnaud \&
Raymond (1992).}
\label{fig:FeXVIIIrates}
\end{figure}

To obtain a total Fe XVIII to Fe XVII $\Delta n=0$ DR rate coefficient,
we have convolved all our measured resonance strengths and energies,
including also the fine-structure excitation channel, with an isotropic
Maxwellian electron distribution (Figure \ref{fig:FeXVIIIrates}).  The
estimated total experimental uncertainty is less than 20\%.  The
published theoretical DR rates are also shown in Figure
\ref{fig:FeXVIIIrates}.  These rates all go rapidly to zero for $k_BT_e
\lesssim 30$ eV because they have not accounted for DR via $2p_{1/2}
\rightarrow 2p_{3/2}$ core excitations.  The Burgess formula (1965)
also does not account for this channel and goes rapidly to zero for low
$k_BT_e$ because the formula is valid only for core excitations
connected to the ground state via an electric dipole transition.  To
sum up, at temperatures of $k_BT_e \sim 15$ eV, near where the
fractional abundance of Fe XVIII is predicted to peak in a photoionized
plasma of cosmic abundances (\cite{Kall96a}), our measured DR rate is a
factor of $\sim 2-200$ times larger than these existing theoretical
rates.

Also shown in Figure \ref{fig:FeXVIIIrates} is the recommended RR rate
of Arnaud \& Raymond (1992).  Using existing theoretical DR rates,
the total recombination rate (RR+DR) at $k_BT_e \sim 15$ eV barely
exceeds the RR rate alone.  Using our inferred DR rates yields a total
recombination rate at $k_BT_e \sim 15$ eV which is a factor of
$\sim 1.5$ larger than the RR rate alone.

For plasma modeling, we have fit our inferred Fe XVIII $\Delta n=0$
Maxwellian-averaged DR rate coefficient to the simple fitting formula
(\cite{Arna92a})
\begin{equation} 
\alpha_{DR}(T_e)=T_e^{-3/2}\sum_i c_i e^{-E_i/k_BT_e}.  
\label{eq:drratefit} 
\end{equation} 
Here $c_i$ and $E_i$ are, respectively, the strength and energy
parameters for the $i$th fitting component.  Best fit values are listed
in Table \ref{tab:fitparameters}.  The fit is good to better than 1.2\%
for $0.05 \le k_BT_e \le 10000$ eV.  Below 0.05 eV, the fit goes to
zero faster than our measured rate.  This is unimportant as RR for
$k_BT_e \le 0.05$ eV is $\gtrsim 600$ times larger than DR.

Contributions due to DR into $n\ge n_{max}=124$, which are not
accessible in our setup are calculated theoretically to
increase the DR rate by $\lesssim 2\%$.  Hence, the zero density DR
rate ($n_{max}=\infty$) is estimated to be $\lesssim 2\%$ larger than
our inferred DR rate.  Our calculations also show that  DR into
$n\ge50(100)$ levels accounts for $\sim 20(10)$\% of the total rate.

\subsubsection{Fe XIX}

The lowest energy, resolved DR resonance lies at $0.0660 \pm0.0005$
eV.  Below $E \sim 0.02 {\rm eV} \lesssim k_BT_\perp$, it is not
possible to resolve resonances from the near 0 eV RR signal.  We can,
however, infer the presence of resonances below $\sim 0.02$ eV.  The
measured Fe XIX recombination rate at $\lesssim 10^{-3}$ eV is over a
factor of $\sim 10$ larger than predicted using semiclassical RR theory
with quantum mechanical corrections (\cite{Schi98a}).  For Fe XVIII,
this rate is only a factor of $\sim 3$ larger.  A number of issues
pertaining to RR measurements at collision energies $\lesssim 10^{-3}$
eV in electron coolers remain to be resolved (\cite{Hoff98a};
\cite{Schi98a}), but it is highly unlikely that their resolution will
lead to RR rates that scale by a factor of $\sim 3$ for a change in
ionic charge from 17 to 18.  Thus, we infer that there are unresolved
DR resonances contributing to the recombination signal below 0.02 eV.
Our calculations suggest they are $2s^2 2p^4(^3P_1)20d$ resonances, but
due to existing experimental and theoretical limitations, it is not
possible unambiguously to identify these resonances.

\begin{figure}
\plotone{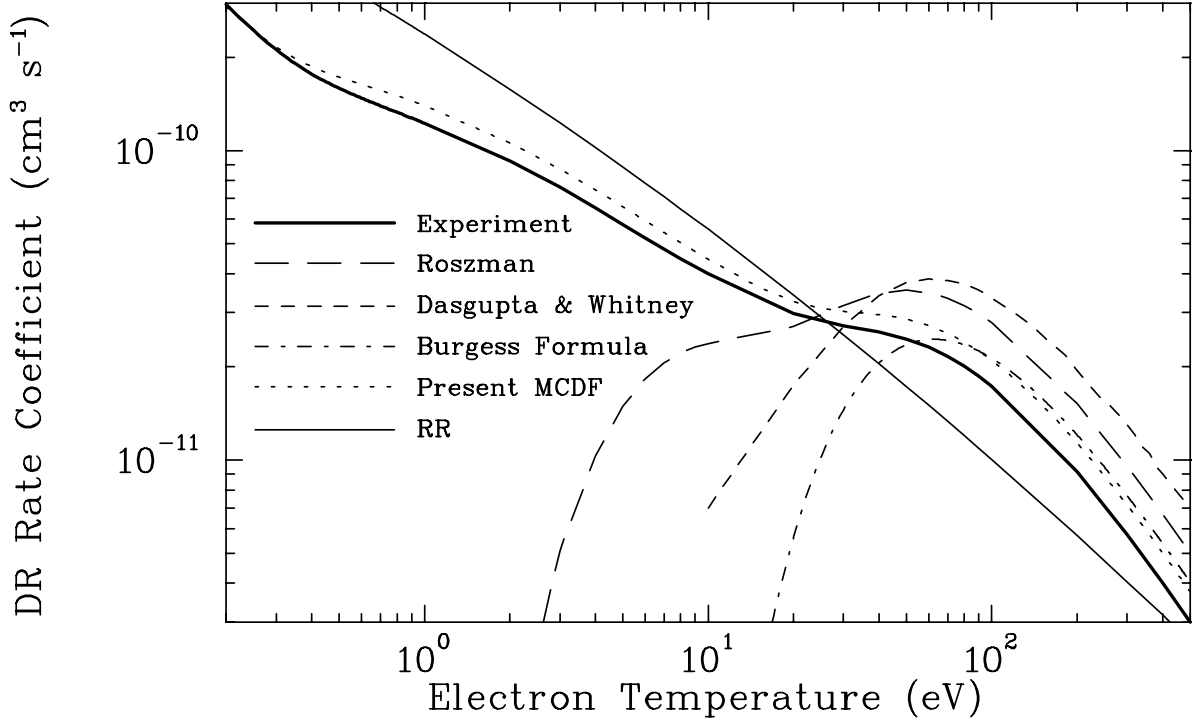}
\caption{Fe XIX to Fe XVIII Maxwellian-averaged $\Delta n=0$ DR rate
coefficients.  The thick solid line is calculated using the measured DR
resonance strengths and energies extracted from the results shown in
Figure \protect\ref{fig:FeXIXresonances} and listed in Table
\protect\ref{tab:FeXIXtable1}.  There is an estimated $\lesssim 20\%$
total systematic uncertainty in our experimentally inferred rate.  Also
shown are existing theoretical calculations by Roszman (1987b;
long-dashed curve) and Dasgupta \& Whitney (1994; short-dashed curve),
the Burgess formula (1965; short-dashed-dot curve) and our new MCDF
calculations (dotted curve).  Our MCBP rate (not shown here) agrees
well with our MCDF rate.  The thin solid curve shows the recommended RR
rate of Arnaud \& Raymond (1992).}
\label{fig:FeXIXrates}
\end{figure}

We have used our measured resonance strengths and energies to calculate
the Fe XIX to Fe XVIII $\Delta n=0$ DR rate coefficient for an
isotropic Maxwellian plasma.  Our calculated rate is shown in Figure
\ref{fig:FeXIXrates} for $k_BT_e \ge 0.2$ eV.  Since the inferred,
unresolved DR resonances below $0.02$ eV are not included in our
derived Maxwellian-averaged rate coefficient, the experimental DR rate 
should go to zero faster than the true DR rate.  As it is
extremely unlikely that Fe XIX will ever form at $k_BT_e \lesssim 0.2$ eV
(\cite{Kall96a}), this uncertainty is expected to have an insignificant
effect on plasma modeling.  Above 0.2 eV we estimate the uncertainty in
the absolute magnitude of our inferred rate to be less than 20\%.

Existing theoretical Fe XIX $\Delta n=0$ rate coefficients are also
shown in Figure \ref{fig:FeXIXrates}.  For $T_e \lesssim 30$ eV the
calculated rates of Roszman (1987b) and Dasgupta \& Whitney (1994) both
underestimate the DR rate, as does the Burgess formula (1965) using the
oscillator strengths of Fuhr et al.\ (1988).  All these calculations
considered only DR via $2s-2p$ core excitations and thus do not include
DR via fine structure core excitations which, as shown by Savin et
al.\ (1997), can be very important.  Below $T_e \sim 30$ eV the rate of
Dasgupta \& Whitney goes to zero faster than that of Roszman.  A
partial explanation is that Dasgupta \& Whitney do not account for DR
into the $n=6$ level.    The exact reason, however, is unclear.
Roszman does not state the $n$ level for which he calculates $2s-2p$ DR
to be energetically allowed.  This level may have been $n=6$; or if
$n=7$, then the calculated resonance energies may be shifted by several
or more eV below the true energies.  For $T_e \gtrsim 30$ eV, both Roszman
and Dasgupta \& Whitney overestimate the DR rate.  This may be partly
due to their not accounting for autoionizations which leave the initial
ion in a $2s^2 2p^4\ ^3P_0$ or $^3P_1$ state.  At temperatures of $\sim
70$ eV, near where Fe XIX is predicted to form in photoionized gas of
cosmic abundances (\cite{Kall96a}), Roszman overestimates the DR rate
by a factor of $\sim 1.5$ and Dasgupta \& Whitney by $\sim 1.7$, and
the Burgess formula by $\sim 1.1.$

Also shown in Figure \ref{fig:FeXIXrates} is the recommended RR rate
of Arnaud \& Raymond (1992).  At $k_BT_e \sim 70$ eV DR dominates
over RR by a factor of $\sim 2$.  Using the recommended RR rate and our
inferred DR rate yields a total recombination rate $\sim 1.4$ smaller
than that obtained using the published DR calculations of Roszman
(1987b) or Dasgupta \& Whitney (1994).

We have fit our inferred, Maxwellian-averaged DR rate coefficient using
Equation \ref{eq:drratefit}.  Best fit parameters are listed in Table
\ref{tab:fitparameters}.  The fit reproduces our rate to better than
4\% for $0.004 \le k_BT_e \le 10000$ eV.  Below 0.004 eV, the fit goes
to zero faster than our measured rate.  
However, for $k_BT_e \lesssim 0.2$ eV, the true rate is likely to
be larger than either the fit or our inferred rate because of additional DR
resonance contributions at $E \lesssim 0.02$ eV.

Contributions due to DR into $n\ge n_{max}=130$, which are not accessible
in our setup, are estimated to increase the DR rate by $\lesssim 4\%$.
Hence, the zero density DR rate ($n_{max}=\infty$) is estimated
theoretically to be $\lesssim 4\%$ larger than our inferred DR rate.
Our calculations also show for Fe XIX that DR into $n\ge50(100)$ levels
accounts for $\sim 20(10)$\% of the total rate.

\section{New Theoretical Calculations}
\label{sec:Theory}

Accurate low temperature DR calculations are challenging both theoretically
and computationally.  Resonance energies often need to be known to
better than $0.01 - 0.10$ eV, which for multi-electron ions can push
theoretical techniques beyond their present capabilities (cf., DeWitt
et al.\ 1996; Schippers et al.\ 1998) Also, approximations must be made
to make the calculations tractable (\cite{Hahn93a}).  To help benchmark
current theoretical capabilities, we have carried out detailed
state-of-the-art MCDF and multiconfiguration Breit Pauli (MCBP)
calculations for comparison with our experimental results.

\subsection{Multiconfiguration Dirac-Fock (MCDF) Method}

DR resonance strengths and rate coefficients for Fe XVIII and Fe XIX
are calculated in the independent processes and isolated resonance
approximation (Seaton \& Storey 1976).  Required transition energies,
Auger and radiative rates are evaluated using the MCDF method in
intermediate coupling with configuration interaction within the same
$n$ complex (Chen 1985; Grant et al. 1980). All possible Coster-Kronig
transitions and radiative transitions to bound states are included.
For $2s+e^- \rightarrow 2pnl$ DR, a one-step cascade stabilization
correction is taken into account when the intermediate state
radiatively decays to another autoionizing state.  All possible
autoionization channels for the recombining ion are accounted for,
including autoionization to an excited state of the initial ion.

For Fe XVIII, we include excitation from the ground state $1s^2 2s^2
2p^5\ ^2P_{3/2}$ to the $1s^2 2s^2 2p^5\ ^2P_{1/2}$ and $1s^2 2s
2p^6\ ^2S_{1/2}$ states.  Explicit calculations are carried out for
$18\le n\le36$ and $l\le8$ for $2p_{1/2}-2p_{3/2}$ core excitations and
for $6\le n\le36$ and $l\le8$ for $2s-2p$ core excitations.  Calculated
excitation energies agree well with measurements (Corliss \& Sugar
1982), and theoretical resonance energies are used without
adjustment.

For Fe XIX, we include excitation from the ground state $1s^2 2s^2
2p^4\ ^3P_2$ to the $1s^2 2s^2 2p^4\ ^3P_0$, $^3P_1$, $^1D_2$, $^1S_0$
and the $1s^2 2s 2p^5\ ^3P^o_{0,1,2}$ and $^1P^o_1$ excited states.  Using
experimental core excitation energies (Corliss \& Sugar 1982), the resonance
energies are adjusted by $\lesssim 1$ eV for all levels except the
$2s^2 2p^4 (^1S_0)$.  The calculated energy of this level is 3.7 eV
larger than its known value because the $2p^6 (^1S_0)$ state is not
included in the configuration-interaction (CI) basis set.  Had it been
included, the calculated energy of the $2s^2 2p^4 (^1S_0)$ state would
have decreased by $\sim 3$ eV.  Its omission from the CI basis set has
an insignificant effect on calculated autoionization rates.  Explicit
calculations are performed for $11\le n \le30$ and $l\le12$ for the
fine-structure transitions (i.e., excitations to the first four excited
states) and for $6\le n \le 30$ and $l \le 12$ for $2s-2p$
excitations.

Extrapolation to higher $n$ Rydberg states for both ions is done by
using an $n^{-3}$ scaling for the Auger and radiative rates. For Fe
XVIII, extrapolations for $l>8$ are calculated using a power law fitted
to $l=6$, 7, and 8.  For Fe XIX, no high-$l$ extrapolation is performed
since by $l=12$ the cross section has already converged to better than
1\%.

\subsection{Multiconfiguration Breit-Pauli (MCBP) Method}

Again, the DR cross sections are calculated in the independent
processes and isolated resonance approximations. Energy levels,
autoionization and radiative rates are calculated in intermediate
coupling using the multi-configuration code AUTOSTRUCTURE (Badnell
1986, 1997). All possible autoionizing transitions and radiative
transitions to bound states are included.  This includes autoionization
of the recombining ion to all energetically allowed states of the
initial ion.  In addition, a one-step cascade is taken account of when
the core electron of the intermediate state radiates and leaves the ion
in an autoionizing state.

For Fe XVIII, explicit calculations are carried out for $6\le n \le
124$ and $0 \le l \le 17$.  For Fe XIX, explicit calculations are
carried-out for $6\le n\le 130$ and $0 \le l \le 15$.  Configuration
mixing within and between $n$ manifolds is taken into account between
all recombined and recombining configurations with $n \le 6$.  For $n
\ge 7$, configuration mixing is restricted to the core only.
Rydberg--Rydberg radiative transitions $n \rightarrow n^\prime$ were
calculated hydrogenically for $n^\prime \ge 7$. The calculated core
energies for Fe XVIII and FeXIX were adjusted by $\lesssim 0.6$ eV to
match the observed values (Kelly 1987).  This gives a marked
improvement to resonance positions, which in general are not
known {\it a priori}.

\subsection{Comparison with Experiment}

Tables \ref{tab:FeXVIIItable1} and \ref{tab:FeXVIIItable2} list the new
theoretical resonance energies and strengths for Fe XVIII DR via the
$^2P_{3/2} -\ ^2P_{1/2}$ and $^2P_{3/2} -\ ^2S_{1/2}$ core excitations,
respectively.  The new theoretical resonance strengths and energies for
Fe XIX are listed in Table \ref{tab:FeXIXtable1}.

\begin{figure}
\plotone{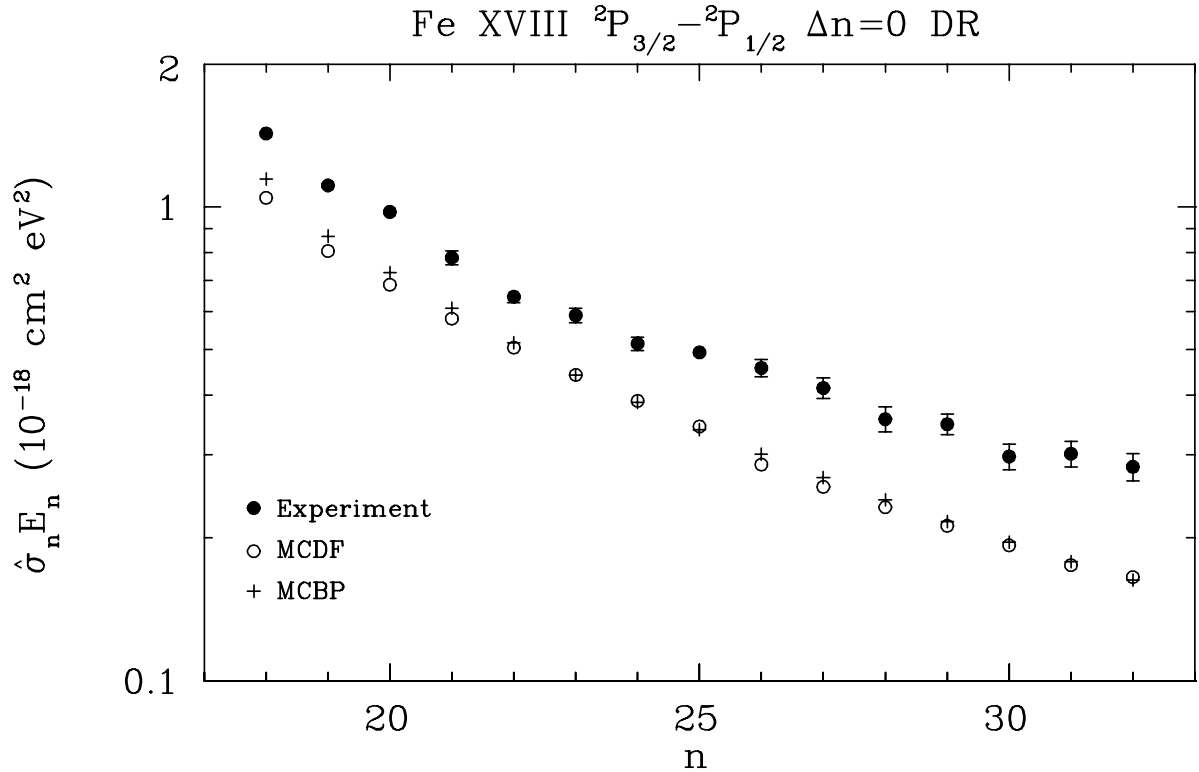}
\caption{DR resonance strength $\hat{\sigma}_n E_n$ as a function of
the principal quantum number $n$ for Fe XVIII to Fe XVII $\Delta n=0$
DR via the $^2P_{3/2}-\ ^2P_{1/2}$ core excitation.  There is an
estimated $\lesssim 20\%$ total systematic uncertainty in our
experimental values (at a 90\% confidence level).  Filled circles are
the present experimental results.  Error bars represent the 1$\sigma$
statistical fitting uncertainties.  Open circles are our MCDF
calculations and crosses are our MCBP calculations.}
\label{fig:FeXVIIIseriesa}
\end{figure}

\begin{figure}
\plotone{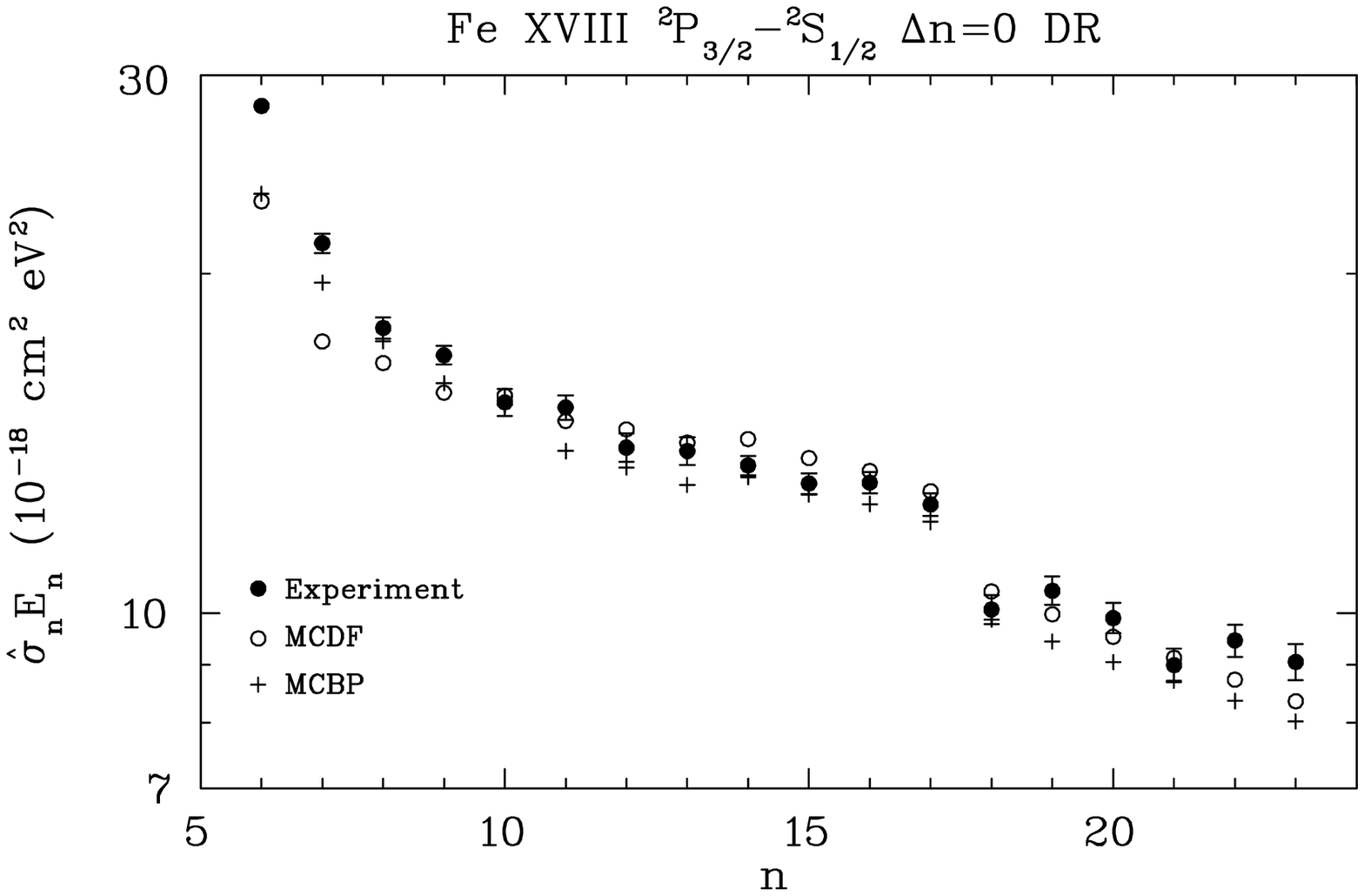}
\caption{DR resonance strength $\hat{\sigma}_n E_n$ as a function of
the principal quantum number $n$ for Fe XVIII to Fe XVII $\Delta n=0$
DR via the $^2P_{3/2}-\ ^2S_{1/2}$ core excitation.  See Figure 
\protect\ref{fig:FeXVIIIseriesa} for further details.}
\label{fig:FeXVIIIseriesb}
\end{figure}

In Figures \ref{fig:FeXVIIIseriesa} and \ref{fig:FeXVIIIseriesb} we
plot for the Fe XVIII $^2P_{3/2} -\ ^2P_{1/2}$ and $^2P_{3/2}
-\ ^2S_{1/2}$ channels, respectively, the experimental and theoretical
values of $\hat{\sigma}_n E_n = E_{nl(l\gtrsim3)} \sum_l
\hat{\sigma}_{nl}$.  We have multiplied $\hat{\sigma}_n$ by $E_n$ to
remove the trivial energy dependence on the right-hand-side of Equation
\ref{eq:resonancestrength}.  There is an $\sim 20\%$ to $\sim 40\%$
discrepancy for the entire $^2P_{3/2} -\ ^2P_{1/2}$ series.  This
increases with $n$ and reaches a value of $\sim 50\%$ for the
summed series limit.  These
differences are larger than the total experimental uncertainty.  For
the $^2P_{3/2} -\ ^2S_{1/2}$ series (Figure \ref{fig:FeXVIIIseriesb}),
there is an $\sim 10-18\%$ difference.
Also clearly visible is the opening up of the $2s2p^6(^2S_{1/2})nl
\rightarrow 2s^22p^5(^2P_{1/2}) +e^-$ autoionization channel near
$E=119$ eV, which causes an abrupt decrease in $\hat{\sigma}_n E_n$
between $n=17$ and 18.  This channel was also observed by Lampert et
al.\ (1996) for the isoelectronic ion Se XXVI.

\begin{figure}
\plotone{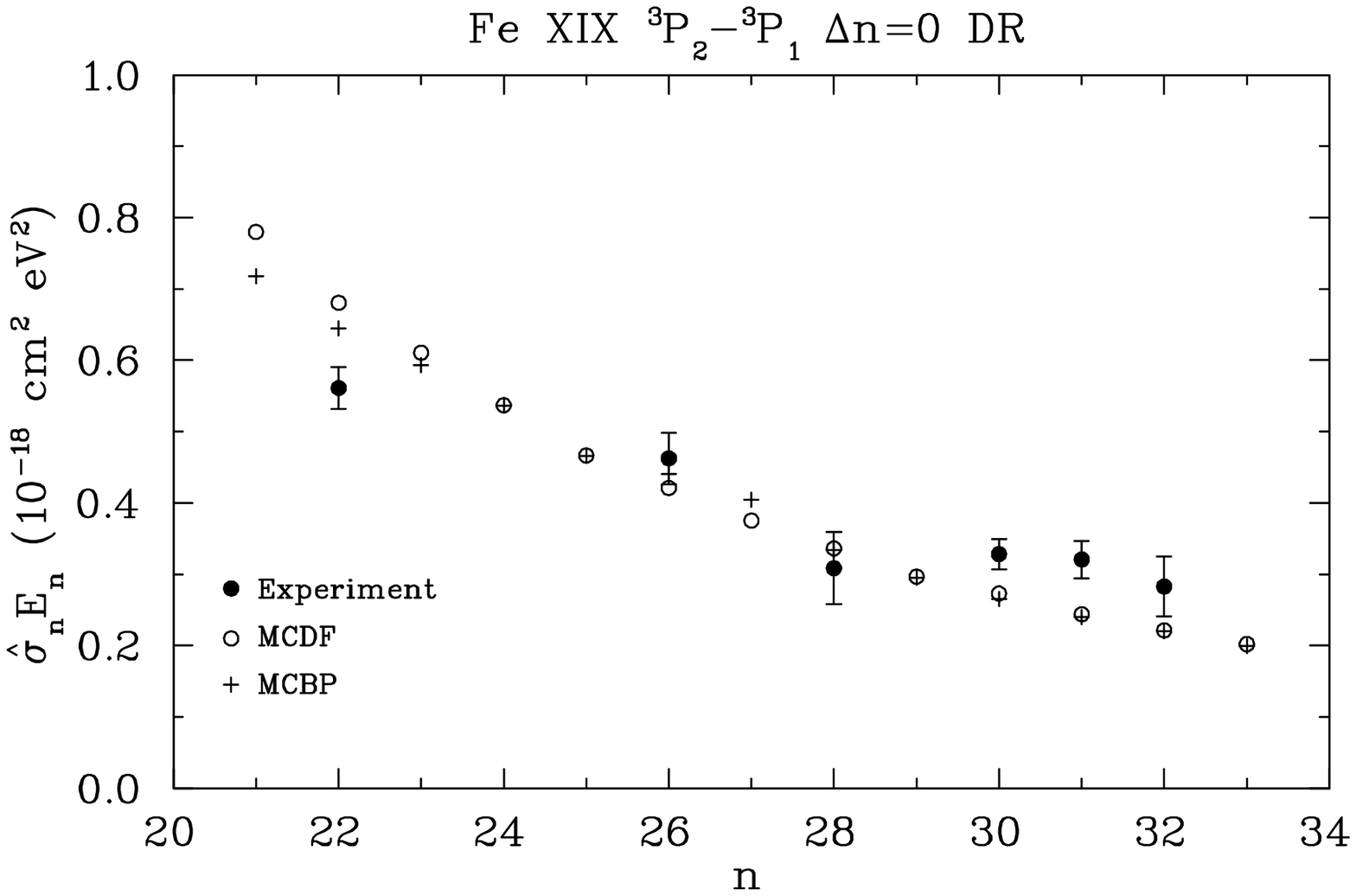}
\caption{DR resonance strength $\hat{\sigma}_n E_n$ as a function of
the principal quantum number $n$ for Fe XIX to Fe XVIII $\Delta n=0$ DR
via the $^3P_2-\ ^3P_1$ core excitation.  See Figure 
\protect\ref{fig:FeXVIIIseriesa} for further details.}
\label{fig:FeXIXseriesa}
\end{figure}

\begin{figure}
\plotone{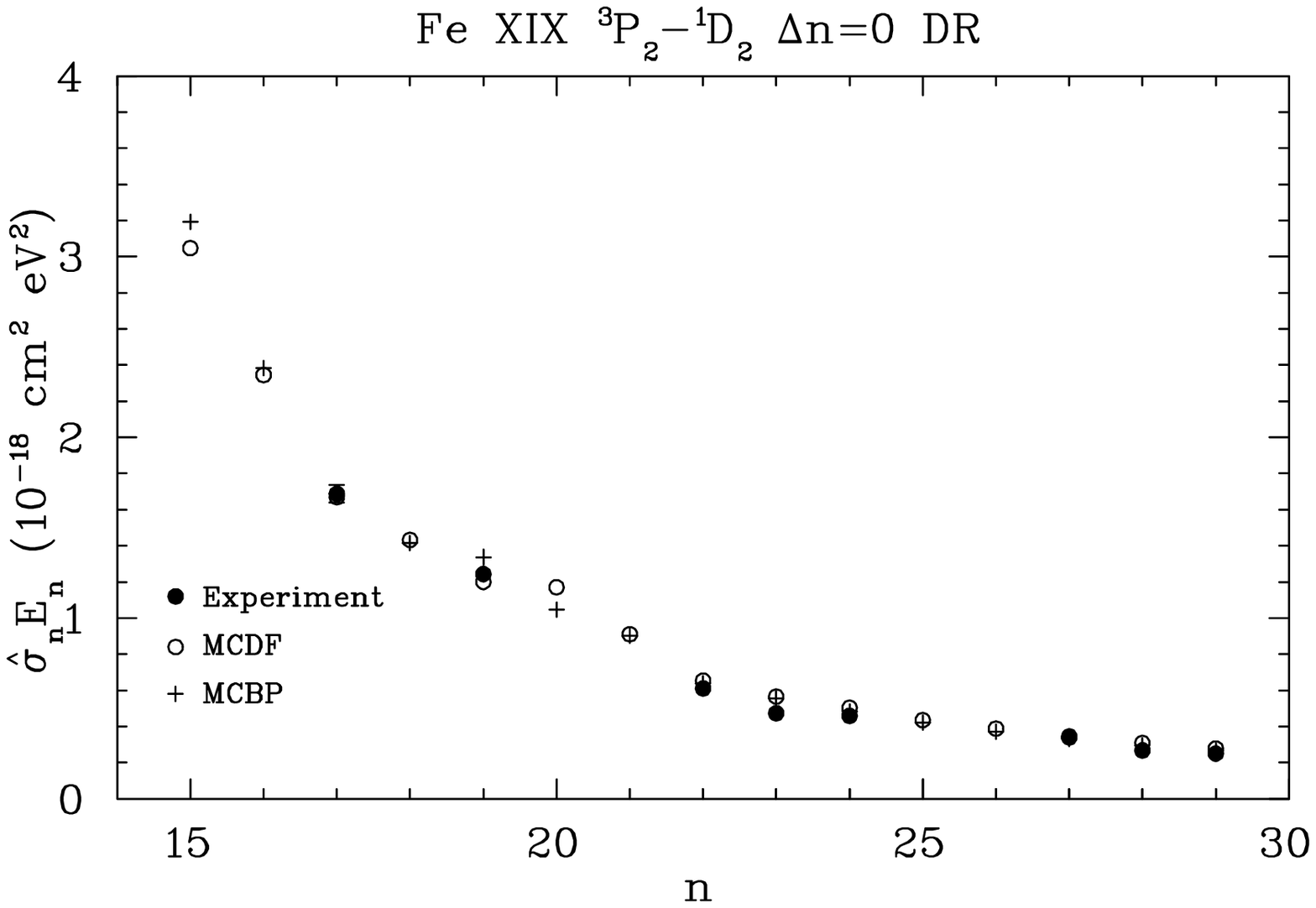}
\caption{DR resonance strength $\hat{\sigma}_n E_n$ as a function of
the principal quantum number $n$ for Fe XIX to Fe XVIII $\Delta n=0$ DR
via the $^3P_2-\ ^1D_2$ core excitation.  See Figure 
\protect\ref{fig:FeXVIIIseriesa} for further details.}
\label{fig:FeXIXseriesb}
\end{figure}

\begin{figure}
\plotone{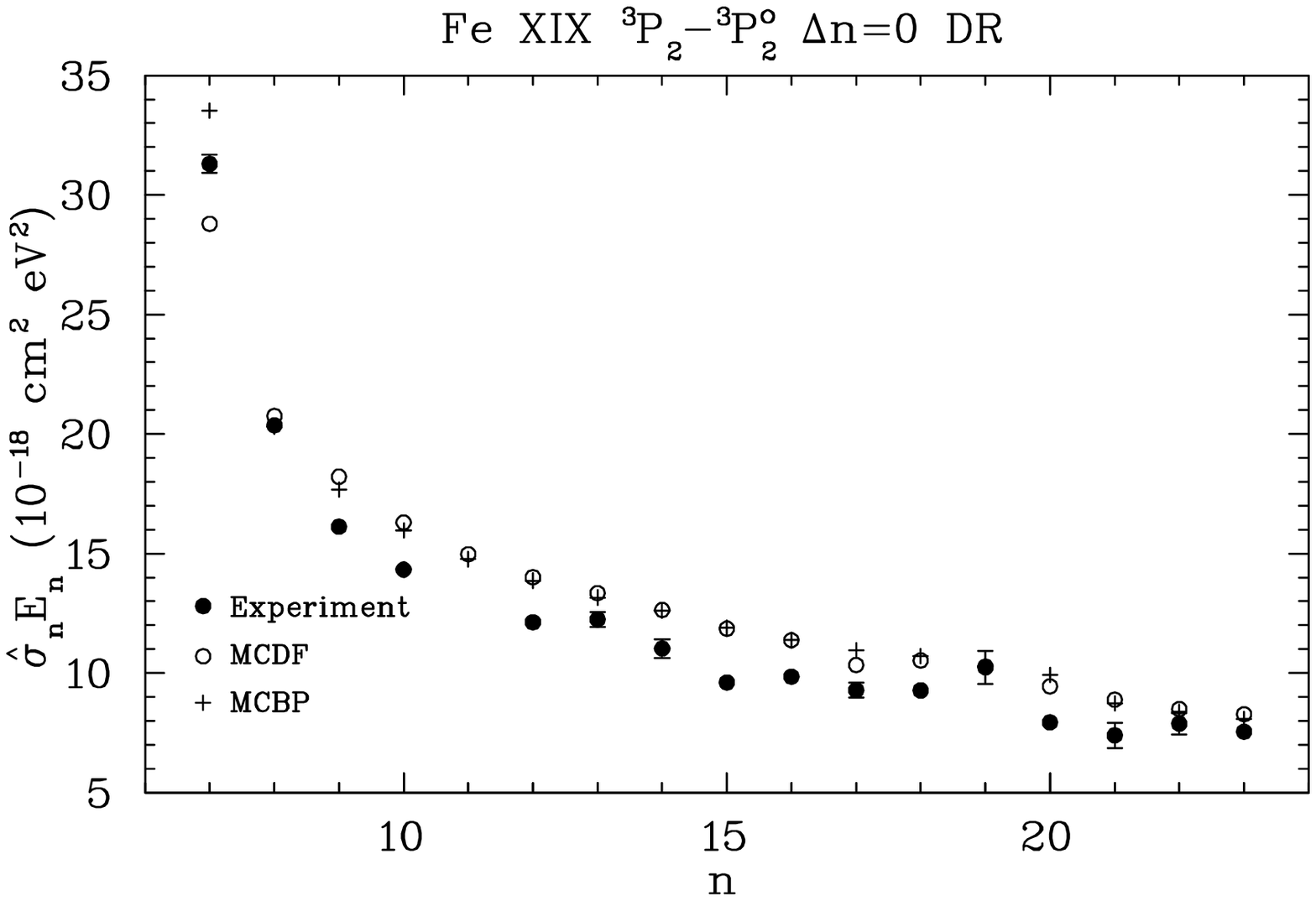}
\caption{DR resonance strength $\hat{\sigma}_n E_n$ as a function of
the principal quantum number $n$ for Fe XIX to Fe XVIII $\Delta n=0$ DR
via the $^3P_2-\ ^3P_2^o$ core excitation.  See Figure 
\protect\ref{fig:FeXVIIIseriesa} for further details.}
\label{fig:FeXIXseriesc}
\end{figure}

\begin{figure}
\plotone{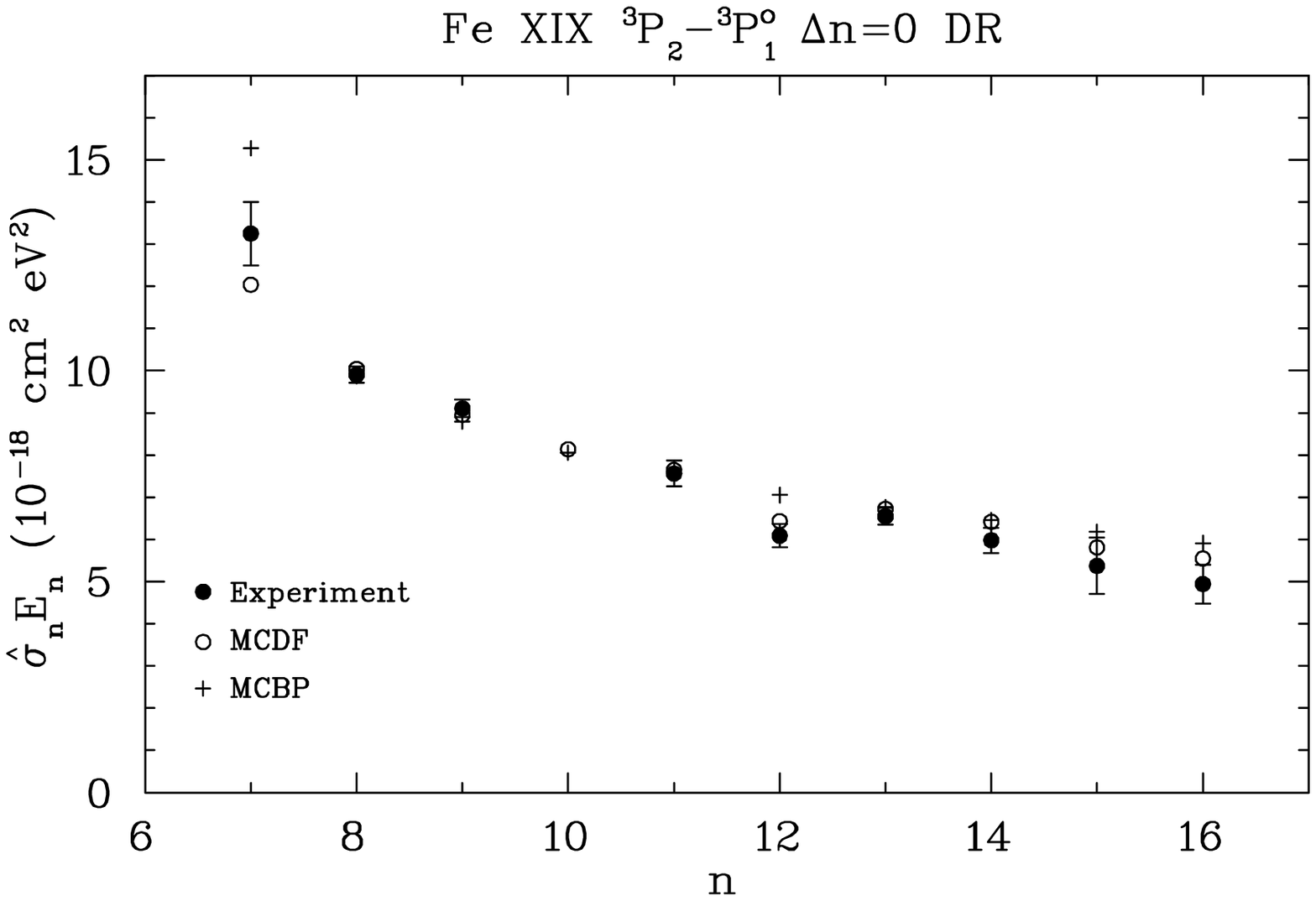}
\caption{DR resonance strength $\hat{\sigma}_n E_n$ as a function of
the principal quantum number $n$ for Fe XIX to Fe XVIII $\Delta n=0$ DR
via the $^3P_2-\ ^3P_1^o$ core excitation.  See Figure 
\protect\ref{fig:FeXVIIIseriesa} for further details.}
\label{fig:FeXIXseriesd}
\end{figure}

In Figures \ref{fig:FeXIXseriesa} to \ref{fig:FeXIXseriesd} we plot
$\hat{\sigma}_n E_n$ using $E_{nl(l\gtrsim4)}$ for the Fe XIX $^3P_2
-\ ^3P_1$, $^3P_2 -\ ^1D_2$, $^3P_2 -\ ^3P^o_2$ and $^3P_2 -\ ^3P^o_1$
DR channels, respectively.  For the $^3P_2 -\ ^3P_1$ and $^3P_2
-\ ^1D_2$ series, we include measured values for only those resonances
which are unambiguously resolved in our experiment from surrounding
resonances.  For example, for the $^3P_2 -\ ^3P_1$ series, we do not
plot the experimental values for $\hat{\sigma}_{29} E_{29}$ because the
$2s^22p^4(^3P_1)29l$ resonance is a weak feature on the shoulder of the
strong $2s^22p^4(^1D_2)17l\ (l\ge2)$ resonance.  However, for the
$^3P_2 -\ ^1D_2$ series we do plot $\hat{\sigma}_{17} E_{17}$ because
any blending from the $2s^22p^4(^3P_1)29l$ resonance is expected to
introduce only a small error.  For this series we also do not plot the
measured value of $\hat{\sigma}_{21} E_{21}$.  It blends with the
$^3P_2 -\ ^3P_1$ series limit.  For the $^3P_2 - \ ^3P^o_2$ and $^3P_2
- \ ^3P^o_1$ series, we use theory to subtract out the different
$2s2p^5(^1P^o_1)6l$ resonance strengths from the various $n=7$
resonances.  This is estimated to introduce a negligible error.  The
$2s2p^5(^3P^o_2)14p$ and $2s2p^5(^3P^o_1)12l\ (l\ge2)$ blend is
resolved using the theoretical resonance strength for the $14p$
resonance.  In general, agreement between experiment and theory is
good, with a few exceptions such as the MCDF resonance strength for the
$2s^22p^4(^3P_0)22l\ (l\ge3)$ resonance and the MCBP resonance strength
for the $2s 2p^5(^3P^o_1)7f$ resonance.  The reason for the
discrepancies between theory and experiment for these resonances as
well as for the summed resonance strengths of the $^3P_2 - \ ^3P^o_2$
and $^3P_2 - \ ^3P^o_1$ series limits is not understood.

The Maxwellian-averaged rate coefficients from our new Fe XVIII and Fe
XIX MCDF DR calculations are plotted in Figures \ref{fig:FeXVIIIrates}
and \ref{fig:FeXIXrates}.  The theoretical rates agree with our
inferred rates to within $\sim 30\%$.  Though not shown, the MCBP rates
agrees well with the MCDF calculations.  We have fit our MCDF rates
using Equation \ref{eq:drratefit}.  Note that for Fe XIX we do not
include the near 0 eV $2s^2 2p^4[^3P_1]20d$ resonances.  Best fit
values are listed in Table \ref{tab:fitparameters}.  For Fe XVIII, the
fit is good to better than 2\% for $0.06 \le k_BT_e \le 10000$ eV.
Below 0.06 eV, the fit goes to zero faster than theory.  For Fe XIX,
the fit is good to better than 2\% for $0.001 \le k_BT_e \le 10000$
eV.

\section{Astrophysical Implications for Photoionized Gas}
\label{sec:AstrophysicalImplications}

\subsection{Ionization Balance Calculations}

Cosmic plasmas are most commonly modeled using the compiled DR rates of
Aldrovandi \& P\'{e}quignot (1973), Shull \& van Steenberg (1982),
Arnaud \& Rothenflug (1985), and Arnaud \& Raymond (1992).  For
photoionized gases the rates of Nussbaumer \& Storey (1983) are often
used.  And recently Nahar \& Pradhan (1994; 1995) and Nahar (1997) have
calculated the unified electron-ion recombination rates (e.g., RR+DR)
for a number of ions.  But for a few exceptions, these rates have all
been calculated either using $LS$-coupling without accounting for $nlj
\rightarrow nl{j^\prime}$ fine-structure transitions or using the
Burgess formula (\cite{Burg65a}), which neither takes these fine
structure transitions into account nor can account for core excitations
not connected to the ground state via an electric dipole transition.

Our results demonstrate that the Burgess formula, $LS$-coupling,
intermediate coupling, and even MCDF calculations can easily under- or
over-estimate the Maxwellian-averaged $\Delta n=0$ DR rate by factors
of $\sim 2$ at ``high'' $T_e$ or underestimate it by factors of $\sim
2$ to orders of magnitude at ``low'' $T_e$.  The limit between ``low''
and ``high'' temperature here is given roughly by the comparison of
$k_BT_e$ with the fine-structure core excitation energy $\Delta E_{fs}$,
typically 10-20 eV for the iron $L$ shell ions.  Our results also
demonstrate that a detailed comparison between theory and experiment of
the resonance strengths and energies that go into the total rate
coefficient is the only way to distinguish unambiguously between
different theoretical rate coefficients.

The importance at some $k_BT_e$ of a given DR channel can be estimated
using Equation \ref{eq:rescond}.  For an ion with fine-structure (i.e.
an ion with a partially filled $p$, $d$, etc., shell), DR via
fine-structure core excitations usually dominates the DR process if the
ion forms at $k_BT_e \lesssim \Delta E_{fs}$.  Nearly all existing
calculations do not account for this channel, and hence they almost
certainly underestimate the $\Delta n=0$ DR rate by factors of $\sim 2$
to orders of magnitude.  For ions which form at $k_BT_e \gtrsim \Delta
E_{fs}$,  fine-structure core excitations are no longer important and
DR is dominated by $nlj \rightarrow nl^\prime{j^\prime}$ ($l \ne
l^\prime$) channels.  Our measurements demonstrate that DR calculations
via these other $\Delta n=0$ channels can readily under- or
over-estimate the DR rate by factors of $\sim 2$.  Taken together, our
results call into question all existing theoretical $\Delta n=0$ DR
rates used for ionization balance calculations of cosmic plasmas.

\subsection{Thermal Instability}

Hess, Kahn, \& Liedahl (1997) showed that Fe $L$ ions play an important
role in determining the range in parameter space over which
photoionized gas is predicted to be thermally unstable.  But they found
the existence of the instability was robust to changes in elemental
abundance and the shape of the ionizing spectrum.  Reynolds
\& Fabian (1995) found the instability was robust to changes in
density, optical depth, and the shape of the ionizing spectrum.  Hess
et al.\ also studied the effects of new Fe $L$ $\Delta n=0$ DR rates
published after the compilation of Arnaud \& Rothenflug (1992) and
found no significant effects.

Our measurements, which demonstrate that published Fe $L$ $\Delta n=0$
DR rates can be wrong by factors of $\sim 2$ or more, call into
question this last conclusion of Hess et al.\ (1997).  We have used
XSTAR (version 1.40b; \cite{Kall97a}) to re-investigate the effects on
the thermal instability of photoionized gas due to our estimated
factor of 2 errors in the Fe XX through Fe XXIV $\Delta n=0$ DR
rates.  For Fe XVIII and Fe XIX, we use our inferred DR rates.  Because
the DR rates in XSTAR do not account for DR via fine-structure core
excitations, we have used the Fe XVIII and Fe XIX results to estimate
the DR rates via $2p_{1/2} \rightarrow 2p_{3/2}$ core excitations for
Fe XX through Fe XXII.  We have run XSTAR using Fe XX through Fe XXIV
$\Delta n=0$ DR rates unchanged, increased by a factor of two, and
decreased by the same factor.  We assume cosmic abundances; and similar
to Reynolds \& Fabian (1995), we assume a model AGN ionizing continuum
consisting of a photon number power law $N\propto E^{-1.8}$ which
extends from 13.6 eV to 40 keV.  Here $E$ is the photon energy.

\begin{figure} 
\plotone{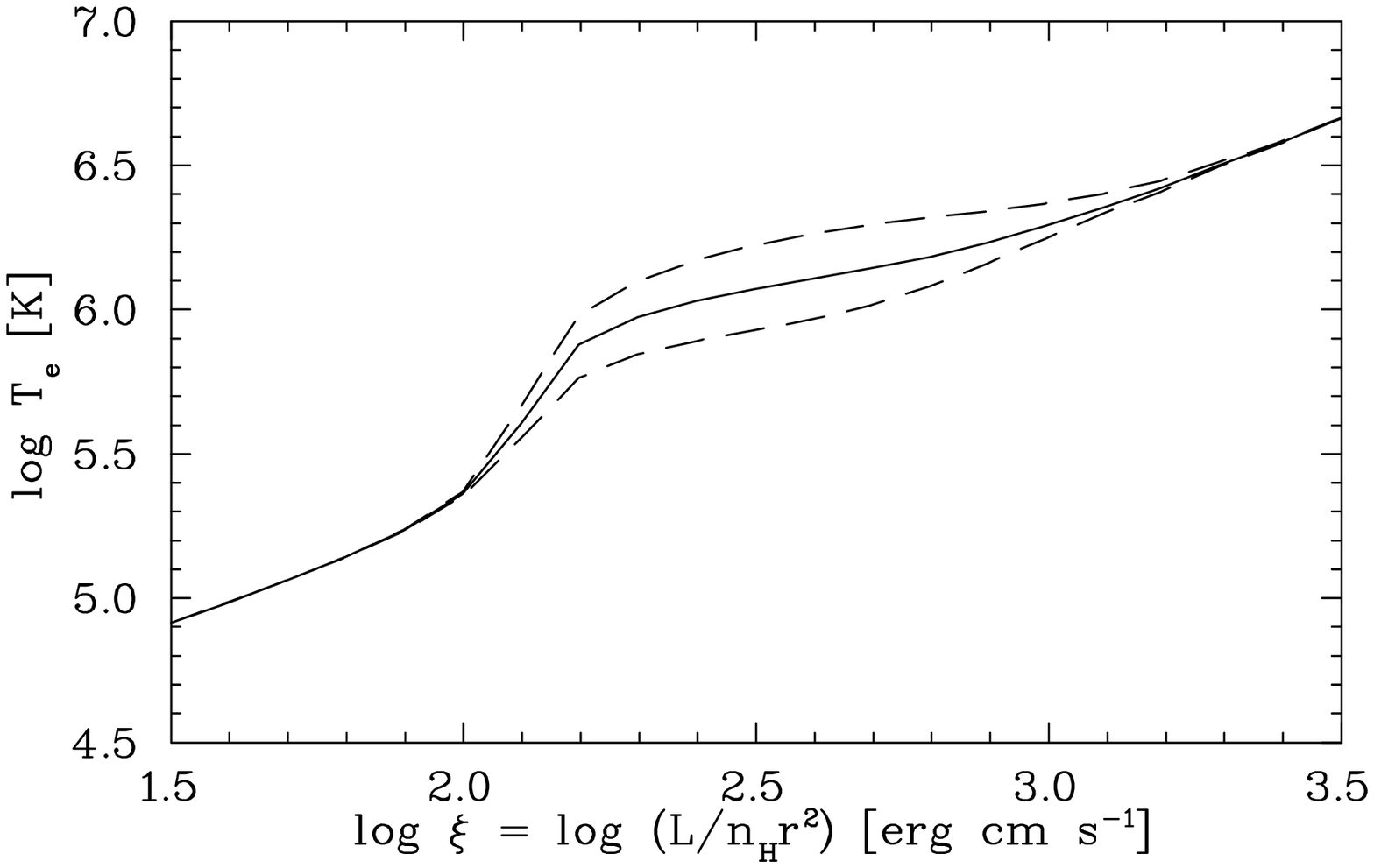}
\caption{Predicted electron temperature versus ionization parameter
$\xi$ for a model AGN ionizing spectrum illuminating a slab
of gas with cosmic abundances.  The solid curve shows the predicted
$T_e$ using our inferred Fe XVIII and Fe XIX DR rates and the unchanged
$\Delta n=0$ DR rates for Fe XX through Fe XXIV.  The upper(lower)
dashed curve results when the $\Delta n=0$ DR rates for Fe XX
through Fe XXIV are increased(decreased) by a factor of 2.}
\label{temperature} 
\end{figure}

\begin{figure} 
\plotone{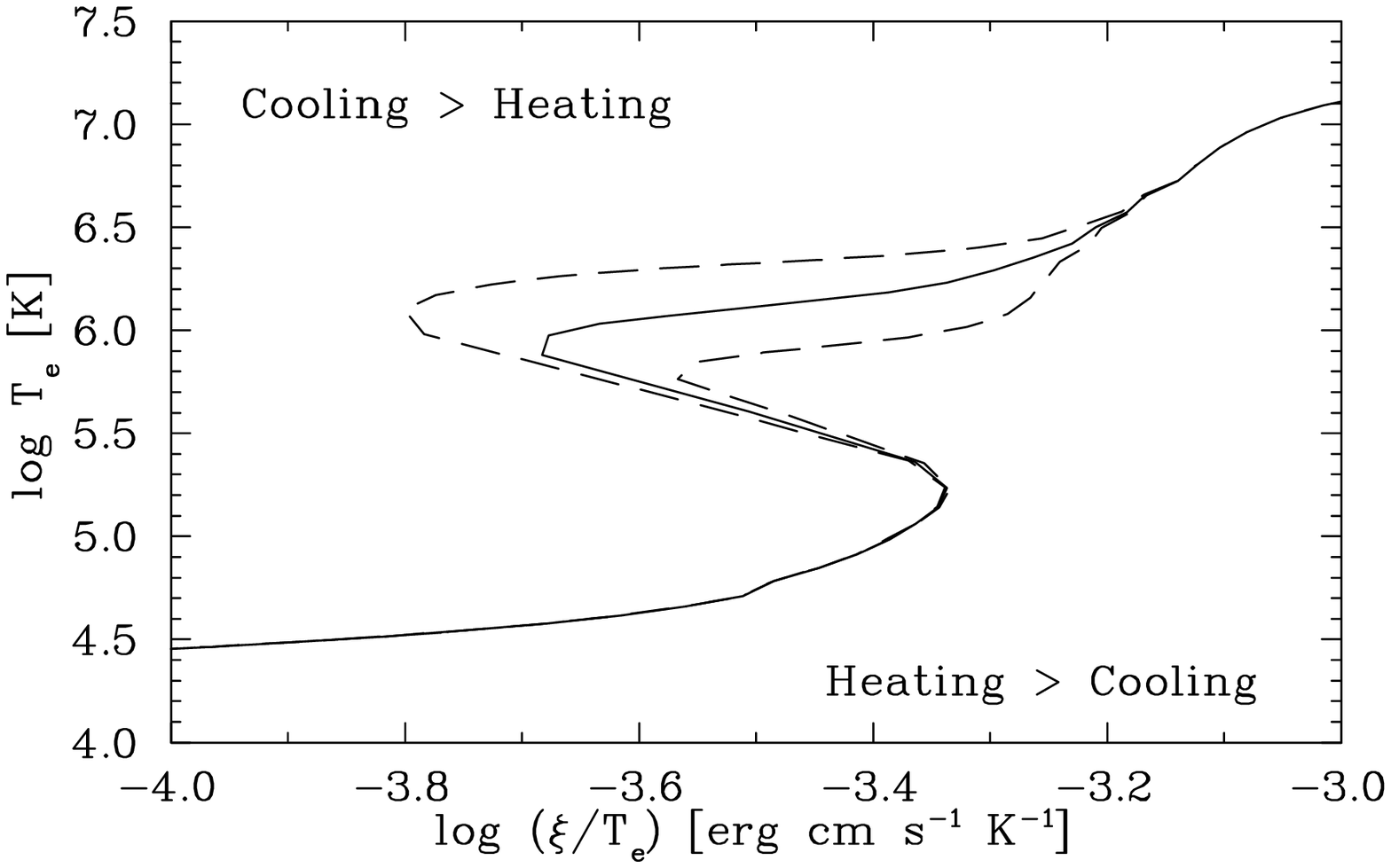}
\caption{Predicted electron temperature versus 
$\xi/T_e$ for a model AGN ionizing spectrum illuminating a slab
of gas with cosmic abundances.  The solid curve shows the predicted
$T_e$ using our inferred Fe XVIII and Fe XIX DR rates and the unchanged
$\Delta n=0$ DR rates for Fe XX through Fe XXIV.  The upper(lower)
dashed curve results when the $\Delta n=0$ DR rates for Fe XX
through Fe XXIV are increased(decreased) by a factor of 2.}
\label{phase} 
\end{figure}

Figure \ref{temperature} shows the predicted $T_e$ versus the
ionization parameter $\xi=L/n_Hr^2$, where $L$ is the luminosity of the
ionizing source, $n_H$ is the hydrogen nucleus density, and $r$ is the
distance from the ionizing source.  Figure \ref{phase} is a phase
diagram of the gas.  For the different DR rates, $T_e$ is shown for
steady-state condition (where heating and cooling of the gas are equal)
versus $\xi/T_e \propto F/p$.  Here $F$ is the ionizing flux and $p$ is
the pressure of the gas.  The well-known thermal instability of
photoionized gas in steady-state can be seen for $-3.80 \lesssim
\log(\xi/T_e) \lesssim -3.35$, where $\xi$ is in units of ergs cm
s$^{-1}$ and $T_e$ in K.

The estimated uncertainty in the DR rates results in as much as a
factor of $\sim 1.8$ difference between predicted values of $T_e$.  And
should a future observation yield $\log T_e \sim 6.1$ (where $T_e$ is
in K), then the uncertainty in the inferred $\xi$ would be a factor of $\sim 3.4$.  Astrophysically, for observations in this range of $\xi$,
these uncertainties will hamper our ability to determine $L$ or $n_e$
to within a factor of $\sim 3.4$ or $r$ to within a factor of $\sim
1.8$.  Also, while the uncertainties in the DR rates do not remove the
thermal instability, they do dramatically affect the range in parameter
space over which the instability is predicted to exist.  The range
changes by a factor of $\sim 1.8$ in $\xi/T_e$ and a factor of $\sim
2.2$ in $T_e$.  When we have completed our measurements for all the Fe
$L$ $\Delta n=0$ DR rates, we will be able to resolve this problem
formally.

The above results demonstrate the effects of the uncertainties in the
DR rates for Fe $L$ ions.  Calculated rates for other ions are likely
to have similar errors.  In order to model photoionized gases
accurately, corrections to the DR rates for all the relevant ions will
be required.  However, the ionization structure of photoionized gas
is not a simple function of temperature.  The temperature at which an
ion forms depends upon the shape of the ionizing spectrum, the
metallicity of the gas, additional heating and cooling mechanisms, and
radiative transfer effects.  An ion forming at a given $T_e$ in one
object could potentially form at a different $T_e$ in another.  Because
it is unknown {\it a priori} what $T_e$ of the observed gas will be, it
is important to use DR rates with the correct $T_e$ dependence over the
entire $T_e$ range of interest.

\subsection{Line Emission}

In photoionized gases, lines produced by $\Delta n=0$ DR provide the
basis for new classes of electron temperature and density diagnostics
(\cite{Lied92a}; \cite{Kahn95a}; \cite{Savi98a}).  DR is a resonance
process and has a $T_e$ dependence different from RR.  Thus, ratios of
DR and RR produced lines can be used as a $T_e$ diagnostic.

One class of $T_e$ diagnostics is based on $2s+e^- \rightarrow 2pnl$ DR
(see \cite{Lied92a} for an extensive discussion; also \cite{Kahn95a}).
For low $n$ values, the recombining ion can radiatively stabilize by a
decay of the captured electron.  This occurs in the presence of an
excited core.  The resulting lines are spectroscopically distinct from
those produced by RR.

Another class of $T_e$ and $n_e$ diagnostics involves DR via
fine-structure core excitations (\cite{Savi98a}).  The excited core
cannot decay via an electric dipole transition and the ion stabilizes by a
radiative decay of the captured electron, which is typically in a high
$n$ level (here, $n \gtrsim 15$).  This leads to an enhancement of $n
\rightarrow 3$ line emission which will appear as broad transition
arrays at {\it AXAF} and {\it XMM} resolution.  Their widths offer a
possible $n_e$ diagnostic.  As $n_e$ increases, the highest $n$ level
which radiatively stabilizes before it is collisionally ionized
decreases.  This reduces the maximum energy of the photons in the
transition array and results in a decrease in the width of the spectral
feature.

A detailed discussion of these various diagnostics will be the topic of
a future paper (Liedahl et al., in preparation).  Further experimental
work is under way to benchmark the DR calculations necessary to develop
these diagnostics.

\acknowledgements

The authors thank T. R. Kallman and S. T. Manson for helpful discussions.
This work was supported in part by NASA High Energy Astrophysics X-Ray
Astronomy Research and Analysis grant NAG5-5123.  Travel and living
expenses for D.W.S. were supported by NATO Collaborative Research Grant
CRG-950911.  The experimental work has been supported in part by the
German Federal Minister for Education, Science, Research, and
Technology (BMBF) under Contract Nos.\ 06 GI 475, 06 GI 848, and 06 HD
854I.  Work performed at Lawrence Livermore National Laboratory was
under the auspices of the US Department of Energy (contract number
W-7405-ENG-48).

\vfill
\clearpage
\eject

\begin{deluxetable}{cclcr}
\footnotesize
\tablecaption{Energy levels (relative to the ground state)
for the $n=2$ shells of Fe XVIII and Fe XIX (\cite{Suga85a})
\label{tab:energylevels}}
\tablewidth{0pt}
\tablehead{\colhead{Ion} & \colhead{ } & \multicolumn{1}{c}{Level} 
& \colhead{ } &  \multicolumn{1}{c}{Energy (eV)}\\
}
\startdata
Fe$^{17+}$ & & $2s^2 2p^5\ ^2P_{3/2}$ & &        0 \nl
           & & $2s^2 2p^5\ ^2P_{1/2}$ & &  12.7182 \nl
           & & $\phn 2s   2p^6\ ^2S_{1/2}$ & & 132.0063 \nl
\nl
Fe$^{18+}$ & & $2s^2 2p^4\ ^3P_2    $ & &        0 \nl
           & & $2s^2 2p^4\ ^3P_0    $ & &   9.3298 \nl
           & & $2s^2 2p^4\ ^3P_1    $ & &  11.0893 \nl
           & & $2s^2 2p^4\ ^1D_2    $ & &  20.9350 \nl
           & & $2s^2 2p^4\ ^1S_0    $ & &  40.3122 \nl
           & & $\phn 2s   2p^5\ ^3P_2^o  $ & & 114.4238 \nl
           & & $\phn 2s   2p^5\ ^3P_1^o  $ & & 122.0922 \nl
           & & $\phn 2s   2p^5\ ^3P_0^o  $ & & 127.7063 \nl
           & & $\phn 2s   2p^5\ ^1P_1^o  $ & & 157.1624 \nl
           & & $\phn \phn \phn      2p^6\ ^1S_0    $ & & 264.6047 \nl
\enddata
\end{deluxetable}

\vfill
\clearpage
\eject

\begin{deluxetable}{cccccccc}
\footnotesize
\tablecaption{Comparison of the measured and calculated Fe XVIII to Fe
XVII $\Delta n=0$ DR resonance energies $E_{nl}$ and energy-integrated
cross sections $\hat{\sigma}_{nl}$ for the $2s^2 2p^5(^2P_{1/2})nl$
resonances.
\label{tab:FeXVIIItable1}}
\tablewidth{0pt}
\tablehead{ 
\colhead{ } & \multicolumn{3}{c}{$E_d$ (eV)} & \colhead{ } &
\multicolumn{3}{c}{$\hat{\sigma}_d$ ($10^{-21}$ cm$^2$ eV)} \\
\cline{2-4} \cline{6-8} \\
\colhead{nl} & \colhead{MCDF\tablenotemark{a}} &
\colhead{MCBP\tablenotemark{a}} &
\colhead{Experiment\tablenotemark{b,c}}  & 
\colhead{ } & \colhead{MCDF} & \colhead{MCBP} &
\colhead{Experiment\tablenotemark{b}} 
}
\startdata
$18s$ & 0.1730 & 0.1981 & $0.2008\pm0.0007$ & & 148.5 & 119.8 & $156.7\pm10.1$ \nl
$18p$ & 0.3087 & 0.3312 & $0.3373\pm0.0007$ & & 200.9 & 253.7 & $278.6\pm 9.1$ \nl
$18d$ & 0.4683 & 0.4908 & $0.4983\pm0.0007$ & & 473.7 & 529.1 & $591.6\pm 8.1$ \nl
$18l$ ($l\ge3$) & 0.5477 & 0.5702 & $0.5818\pm0.0003$ & & 1083.7 & 1105.4 & 
 $1425.9\pm9.3$ \nl
$n=18$ (sum) & & & & & 1906.8 & 2008.4 & $2452.8\pm 18.3$ \nl
$19s$ & 1.475 & 1.499 & $1.494\pm0.008$ & &  13.1 & 13.5 & $ 18.6\pm 5.2$ \nl
$19p$ & 1.590 & 1.612 & $1.621\pm0.003$ & &  36.6 & 44.9 & $ 52.8\pm 5.9$ \nl
$19d$ & 1.725 & 1.748 & $1.760\pm0.002$ & & 109.9 & 127.4 & $139.5\pm 7.7$ \nl
$19l$ ($l\ge3$) & 1.793 & 1.815 & $1.831\pm0.002$ & & 290.1 & 291.8 & $ 394.8\pm 7.8$ \nl
$n=19$ (sum) & & & & & 449.7 & 477.6 & $605.7\pm 13.5$ \nl
$20s$ & 2.583 & 2.607 & $2.631\pm0.015$ & &   6.4 & 6.6 & $  9.7\pm 2.7$ \nl
$20p$ & 2.682 & 2.704 & $2.728\pm0.008$ & &  18.6 & 22.9 & $ 28.9\pm 3.9$ \nl
$20d$ & 2.798 & 2.820 & $2.839\pm0.004$ & &  58.3 & 67.6 & $ 61.6\pm 5.9$ \nl
$20l$ ($l\ge3$) & 2.855 & 2.878 & $2.898\pm0.001$ & & 156.4 & 155.1 & $ 236.3\pm 6.2$ \nl
$n=20$ (sum) & & & & & 239.7 & 252.2 & $336.5\pm 8.6$ \nl
$21s$ & 3.535 & 3.569 & $ 3.572\pm0.022$ & &  4.0 & 3.6 & $  9.7\pm 3.9$ \nl
$21p$ & 3.620 & 3.649 & $ 3.666\pm0.020$ & & 11.8 & 13.2 & $ 11.6\pm 3.9$ \nl
$21l$ ($l\ge2$) & 3.756 & 3.792 & $3.802\pm0.001$ & & 138.7 & 144.3 & $ 184.1\pm 4.1$ \nl
$n=21$ (sum) & & & & & 154.5 & 161.1 & $205.4\pm 6.9$ \nl
$22l$ ($l\ge0$) & 4.536 & 4.560 & $4.591\pm0.002$ & & 111.1 & 113.2 & $ 140.7\pm 4.2$ \nl
$23l$ ($l\ge0$) & 5.231 & 5.254 & $5.281\pm0.002$ & &  84.4 & 83.9 & $ 111.7\pm 4.0$ \nl
$24l$ ($l\ge0$) & 5.841 & 5.864 & $5.890\pm0.002$ & &  66.7 & 66.0 & $  87.3\pm 2.8$ \nl
$25l$ ($l\ge0$) & 6.379 & 6.401 & $6.427\pm0.003$ & &  53.9 & 52.9 & $  76.7\pm 2.7$ \nl
$26l$ ($l\ge0$) & 6.856 & 6.878 & $6.902\pm0.003$ & &  44.8 & 43.7 & $  66.2\pm 2.8$ \nl
$27l$ ($l\ge0$) & 7.281 & 7.303 & $7.324\pm0.004$ & &  37.4 & 36.7 & $  56.5\pm 2.8$ \nl
$28l$ ($l\ge0$) & 7.661 & 7.683 & $7.696\pm0.004$ & &  31.9 & 31.3 & $  46.3\pm 2.8$ \nl
$29l$ ($l\ge0$) & 8.002 & 8.024 & $8.031\pm0.004$ & &  27.7 & 27.0 & $  43.3\pm 2.2$ \nl
$30l$ ($l\ge0$) & 8.310 & 8.332 & $8.349\pm0.005$ & &  24.1 & 23.5 & $  35.6\pm 2.2$ \nl
$31l$ ($l\ge0$) & 8.589 & 8.611 & $8.624\pm0.005$ & &  21.1 & 20.7 & $  34.9\pm 2.2$ \nl
$32l$ ($l\ge0$) & 8.841 & 8.852 & $8.862\pm0.006$ & &  18.7 & 18.4 & $  31.9\pm 2.2$ \nl
$33l\le n \lesssim 124l$ ($l\ge0$) & & & $8.94-12.72$& & 218.0 & 222.7 & 
 $340.6\pm 14.5$ \nl
\enddata
\tablenotetext{a}{Resonance strength weighted energy:
$E_d=\sum_i E_i \sigma_i / \sum_i \sigma_i$.}
\tablenotetext{b}{1$\sigma$ statistical fitting uncertainties only.}
\tablenotetext{c}{Absolute energy scale uncertainty $\lesssim 0.4 \%$.}
\end{deluxetable}

\vfill
\clearpage
\eject

\begin{deluxetable}{cccccccc}
\footnotesize
\tablecaption{Comparison of the measured and calculated Fe XVIII to Fe
XVII $\Delta n=0$ DR resonance energies $E_{nl}$ and energy-integrated
cross sections $\hat{\sigma}_{nl}$ for the $2s 2p^6(^2S_{1/2}) nl$
resonances.
\label{tab:FeXVIIItable2}}
\tablewidth{0pt}
\tablehead{ 
\colhead{ } & \multicolumn{3}{c}{$E_d$ (eV)} & \colhead{ } &
\multicolumn{3}{c}{$\hat{\sigma}_d$ ($10^{-21}$ cm$^2$ eV)} \\
\cline{2-4} \cline{6-8} \\
\colhead{nl} & \colhead{MCDF\tablenotemark{a}} &
\colhead{MCBP\tablenotemark{a}} &
\colhead{Experiment\tablenotemark{b,c}}  & 
\colhead{ } & \colhead{MCDF} & \colhead{MCBP} &
\colhead{Experiment\tablenotemark{b}} 
}
\startdata
$6s\ (J=1)$ & 12.107 & 11.955 & $11.879\pm0.005$ & & 54.2 & 51.6 & $ 55.7\pm 3.4$ \nl
$6s\ (J=0)$ & 12.685 & 12.503 & $12.374\pm0.011$ & & 15.6 & 16.1& $ 25.8\pm 2.9$ \nl
$6p_{1/2}\ (J=0,1)$  & 15.909 & 15.715 & $15.599\pm0.010$ & & 27.3 & 27.9 & $ 30.7\pm 4.7$ \nl
$6p_{3/2}\ (J=1,2)$  & 16.238 & 16.039 & $15.917\pm0.008$ & & 72.1 & 78.8 & $109.2\pm 6.3$ \nl
$6d_{3/2}\ (J=1,2),\ 6d_{5/2}\ (J=3)$ & 20.520 & 20.332 & $20.203\pm0.003$ & 
 & 133.1 & 132.1 & $144.7\pm 4.0$ \nl
$6d_{5/2}\ (J=2)$ & 21.000 & 20.851 & $20.678\pm0.007$ & & 39.6 & 41.8 & $ 49.5\pm 3.8$ \nl
$6f$ & 22.693 & 22.542 & $22.395\pm0.002$ & & 260.2 & 272.3 & $299.0\pm 7.1$ \nl
$6l$ ($l\ge4$) & 22.950 & 22.798 & $22.709\pm0.001$ & & 408.3 & 411.9 & $525.6\pm 7.3$ \nl
$n=6$ (sum) & & & & & 1010.3 & 1032.5 & $1240.2\pm 14.8$ \nl
$7s$ & 45.282 & 45.172 & $44.789-45.938$\tablenotemark{d} & & 9.3 & 12.3 & $ 21.8\pm 4.7$ \nl
$7p$ & 47.702 & 47.533 & $47.438\pm0.017$ & & 23.6 & 35.2 & $ 38.0\pm 3.7$ \nl
$7d$ & 50.474 & 50.309 & $50.175\pm0.011$ & & 53.2 & 60.8 & $ 62.2\pm 3.6$ \nl
$7l$ ($l\ge3$) & 51.882 & 51.683 & $51.647\pm0.003 $ & & 249.6 & 271.8 & $290.0\pm 4.1$ \nl
$n=7$ (sum) & & & & & 335.7 & 380.1 & $412.0\pm 8.1$ \nl
$8s$ & 66.332 & 66.186 & $65.761-66.410$\tablenotemark{d} & & 5.3 & 6.0 & $  4.1\pm 3.0$ \nl
$8p$ & 67.934 & 67.751 & $67.688\pm0.032$ & & 13.3 & 18.1 & $ 12.6\pm 3.1$ \nl
$8d$ & 69.769 & 69.588 & $69.522\pm0.022$ & & 31.2 & 33.5 & $ 29.3\pm 3.2$ \nl
$8l$ ($l\ge3$) & 70.722 & 70.526 & $70.515\pm0.003$ & & 185.8 & 189.4 & $207.9\pm 3.3$ \nl
$n=8$ (sum) & & & & & 235.6 & 247.0 & $253.9\pm 5.5$ \nl
$9p$ & 81.681 & 81.493 & $81.422\pm0.027$ & &  9.5 & 12.4 & $  7.1\pm 1.7$ \nl
$9d$ & 82.959 & 82.772 & $82.675\pm0.019$ & & 22.4 & 23.3 & $ 22.8\pm 2.3$ \nl
$9l$ ($l\ge3$) & 83.631 & 83.437 & $83.404\pm0.003$ & & 155.7 & 156.0 & $173.1\pm 2.7$ \nl
$n=9$ ($l\ge1$ sum) & & & & & 187.6 & 191.6 & $203.0\pm 3.9$ \nl
$10p$ & 91.445 & 91.256 & $90.876-91.476$\tablenotemark{d} & & 7.7 & 9.6 & $  5.5\pm 2.3$ \nl
$10l$ ($l\ge2$) &  92.808 & 92.612 & $ 92.620\pm0.005$ & & 160.2 & 155.6 & $160.6\pm 4.0$ \nl
$n=10$ ($l\ge1$ sum) & & & & & 167.9 & 165.2 & $166.1\pm 4.6$ \nl
$11l$ ($l\ge2$) &  99.651 & 99.459 & $ 99.418\pm0.007$ & & 148.6 & 140.0 & $153.1\pm 3.9$ \nl
$12l$ ($l\ge2$) & 104.855 & 104.663 & $104.658\pm0.007$ & & 138.8 & 128.6 & $134.0\pm 3.9$ \nl
$13l$ ($l\ge2$) & 108.903 & 108.711 & $108.701\pm0.007$ & & 130.1 & 119.6 & $128.1\pm 3.6$ \nl
$14l$ ($l\ge1$) & 112.094 & 111.898 & $111.925\pm0.005$ & & 127.3 & 118.0 & $120.8\pm 2.4$ \nl
$15l$ ($l\ge1$) & 114.678 & 114.493 & $114.525\pm0.007$ & & 119.7 & 111.3 & $113.8\pm 2.4$ \nl
$16l$ ($l\ge0$) & 116.800 & 116.608 & $116.618\pm0.007$ & & 114.5 & 107.2 & $112.0\pm 2.4$ \nl
$17l$ ($l\ge0$) & 118.561 & 118.368 & $118.374\pm0.007$ & & 108.2 & 101.8 & $ 105.5\pm 2.4$ \nl
$18l$ ($l\ge0$) & 120.038 & 119.844 & $119.846\pm0.009$ & &  87.1 & 82.4 & $ 84.1\pm 2.5$ \nl
$19l$ ($l\ge0$) & 121.285 & 121.092 & $121.075\pm0.009$ & &  82.3 & 78.0 & $ 86.5\pm 2.5$ \nl
\enddata
\end{deluxetable}

\setcounter{table}{2}
\begin{deluxetable}{cccccccc}
\footnotesize
\tablecaption{Continued.}
\tablewidth{0pt}
\tablehead{ 
\colhead{ } & \multicolumn{3}{c}{$E_d$ (eV)} & \colhead{ } &
\multicolumn{3}{c}{$\hat{\sigma}_d$ ($10^{-21}$ cm$^2$ eV)} \\
\cline{2-4} \cline{6-8} \\
\colhead{nl} & \colhead{MCDF\tablenotemark{a}} &
\colhead{MCBP\tablenotemark{a}} &
\colhead{Experiment\tablenotemark{b,c}}  & 
\colhead{ } & \colhead{MCDF} & \colhead{MCBP} &
\colhead{Experiment\tablenotemark{b}} 
}
\startdata
$20l$ ($l\ge0$) & 122.350 & 122.157 & $122.130\pm0.010$ & &  77.9 & 74.1 & $ 81.1\pm 2.5$ \nl
$21l$ ($l\ge0$) & 123.267 & 123.074 & $123.051\pm0.012$ & &  74.1 & 70.8 & $ 73.1\pm 2.5$ \nl
$22l$ ($l\ge0$) & 124.061 & 123.869 & $123.872\pm0.013$ & &  70.4 & 67.5 & $ 76.4\pm 2.5$ \nl
$23l$ ($l\ge0$) & 124.754 & 124.562 & $124.577\pm0.017$ & &  67.0 & 64.4 & $ 72.7\pm 2.8$ \nl
$24l \le n \lesssim 124l$ ($l\ge0$) & & & $125.0-132.5$ & & 1575.9 & 1552.3 & $1532.1\pm 8.9$ \nl
\enddata
\tablenotetext{a}{Resonance strength weighted energy:
$E_d=\sum_i E_i \sigma_i / \sum_i \sigma_i$.}
\tablenotetext{b}{1$\sigma$ statistical fitting uncertainties only.}
\tablenotetext{c}{Absolute energy scale uncertainty $\lesssim 0.4 \%$.}
\tablenotetext{d}{Unable to fit for resonance energy.}
\end{deluxetable}

\vfill
\clearpage
\eject

\begin{deluxetable}{ccccccccc}
\footnotesize
\tablecaption{Comparison of the measured and calculated resonance
energies $E_{d}$ and energy-integrated DR cross sections $\hat{\sigma}_d$
of Fe XIX to Fe XVIII $\Delta n=0$ DR.
\label{tab:FeXIXtable1}}
\tablewidth{0pt}
\tablehead{ 
\colhead{ } & \colhead{ } &
\multicolumn{3}{c}{$E_d$ (eV)} & \colhead{ } &
\multicolumn{3}{c}{$\hat{\sigma}_d$ ($10^{-21}$ cm$^2$ eV)} \\
\cline{3-5} \cline{7-9} \\
\colhead{Resonance} & \colhead{ } & 
\colhead{MCDF\tablenotemark{a}} & \colhead{MCBP\tablenotemark{a}} &
\colhead{Experiment\tablenotemark{b,c}}  & 
\colhead{ } & \colhead{MCDF} & \colhead{MCBP} &
\colhead{Experiment\tablenotemark{b}} 
}
\startdata
$2s^22p^4(^3P_1)20l\ (l\ge3)$ & & 0.0619 & 0.0623& & & 6218.7 &7375.5 & \nl
$2s^22p^4(^3P_0)22p$ & & 0.0993 & 0.0970& &  & 186.8 & 226.3& \nl
blend & & 0.0628 & 0.0633 & $0.0660\pm0.0005$ &  & 6405.5 & 7601.8 &
$6726.2\pm180.2$ \nl
& & & & & & \nl
$2s^22p^4(^3P_0)22d$ & & 0.1792 & 0.1754 & $0.1824\pm0.0007$ & &  391.0 & 455.0 &
  $360.4\pm 15.1$ \nl
$2s^22p^4(^3P_0)22l\ (l\ge3)$ & & 0.2236 & 0.2204 & $0.2294\pm0.0004$ & & 770.1 & 658.2 &
  $538.4\pm  9.3$ \nl
$2s^22p^4(^1D_2)15s$ & & 0.7064 & 0.7103 & $0.7054\pm0.0062$ & &   126.5 & 140.1 &
  $110.5\pm  7.0$ \nl
& & & & & & \nl
$2s^22p^4(^1D_2)15p$ & & 0.9410 &0.9450 & & & 260.3 &305.1 & \nl
$2s^22p^4(^3P_1)21p$ & & 0.9421 &0.9456 & & &  68.8 &85.6 & \nl
$2s^22p^4(^3P_0)23d$ & & 0.9601 &0.9562 & & &  64.1 &73.1 & \nl
blend & & 0.9444 & 0.9469 & $0.9545\pm0.0012$ & &   393.2 & 463.8 &
  $429.9\pm  7.3$ \nl
& & & & & & \nl
$2s^22p^4(^3P_1)21l\ (l\ge2)$ & & 1.0677 & 1.0657 & $1.0780\pm0.0017$ & & 661.6 & 587.8 &
  $487.3\pm  7.1$ \nl
$2s^22p^4(^1D_2)15d$ & & 1.2017 & 1.1999 & $1.2052\pm0.0003$ & &   832.4 & 928.4 &
  $746.1\pm  4.8$ \nl
$2s^22p^4(^1D_2)15l\ (l\ge3)$ & & 1.3323 & 1.3299 & $1.3366\pm0.0003$ & & 1067.8 & 1028.1 &
  $871.5\pm 4.7$ \nl
$2s^22p^4(^3P_0)24l\ (l\ge1)$ & & 1.6598 & 1.6552 & $1.6812\pm0.0054$ & &  114.6 & 108.9 &
  $99.4\pm  4.8$ \nl
$2s^22p^4(^3P_1)22p$ & & 1.8502 & 1.8585 & $1.8282\pm0.0157$ & &    30.4 & 37.0 &
  $22.5\pm  7.1$ \nl
$2s^22p^4(^3P_1)22l\ (l\ge2)$ & & 1.9592 & 1.9584 & $1.9783\pm0.0050$ & &  316.9 & 292.3 &
  $261.1\pm 7.9$ \nl
$2s^22p^4(^3P_0)25l\ (l\ge0)$ & & 2.2627 & 2.2568 & $2.2877\pm0.0062$ & &   82.4 & 74.7 &
  $74.0\pm  5.2$ \nl
$2s^22p^4(^3P_1)23l\ (l\ge0)$ & & 2.7235 &2.7102 & & & 224.1 &218.9 & \nl
$2s^22p^4(^3P_0)26l\ (l\ge0)$ & & 2.7971 &2.7940 & & &  59.5 &63.9 & \nl
$2s^22p^4(^1S_0)11p$          & & 2.8421 &2.8374 & & &  47.7 &50.4 & \nl
$2s2p^5(^3P_0^o)6d\ (J=5/2)$  & & 2.8788 &2.9405 & & &  73.2 &75.0 & \nl
blend & & 2.7764 & 2.7813 & $2.7888\pm0.0083$ & &  404.5
& 408.2 & $367.2\pm12.9$ \nl
& & & & & & & & \nl
$2s2p^5(^3P_0^o)6d\ (J=3/2)$ & & 3.0216 & 3.0655 & $2.9699\pm0.0233$ & &   42.2 & 27.7 &
 $44.7\pm 11.4$ \nl
& & & & & & & & \nl
$2s^22p^4(^1D_2)16l\ (l\le1)$ & & 3.3315 &3.3797 & & & 83.2 &95.1 & \nl
$2s^22p^4(^3P_1)24l\ (l\ge0)$ & & 3.4072 &3.4072 & & & 157.6 &157.5 & \nl
$2s^22p^4(^1S_0)11d$ & & 3.4983 &3.4992 & & & 154.4 &163.8 & \nl
blend & & 3.4267 & 3.4292 & $3.4852\pm0.0030$ & &  395.2 & 416.4 & $298.3\pm7.7$ \nl
& & & & & & & & \nl
$2s^22p^4(^1D_2)16l\ (l\ge2)$ & & 3.6627 &3.6560 & & & 556.8 &556.6 & \nl
$2s^22p^4(^1S_0)11f$& & 3.8198 &3.8179 & & & 42.2 &26.9 & \nl
blend & & 3.6739 & 3.6635 & $3.7175\pm0.0019$ & & 599.0 & 583.5 & $574.2\pm8.9$ \nl
\enddata
\end{deluxetable}

\setcounter{table}{3}
\begin{deluxetable}{ccccccccc}
\footnotesize
\tablecaption{Continued.}
\tablewidth{0pt}
\tablehead{ 
\colhead{ } & \colhead{ } &
\multicolumn{3}{c}{$E_d$ (eV)} & \colhead{ } &
\multicolumn{3}{c}{$\hat{\sigma}_d$ ($10^{-21}$ cm$^2$ eV)} \\
\cline{3-5} \cline{7-9} \\
\colhead{Resonance} & \colhead{ } & 
\colhead{MCDF\tablenotemark{a}} & \colhead{MCBP\tablenotemark{a}} &
\colhead{Experiment\tablenotemark{b,c}}  & 
\colhead{ } & \colhead{MCDF} & \colhead{MCBP} &
\colhead{Experiment\tablenotemark{b}} 
}
\startdata
$2s^22p^4(^1S_0)11l\ (l\ge4)$ & & 3.8731 &3.8727 & & & 17.4 &24.4 & \nl
$2s^22p^4(^3P_1)25l\ (l\ge0)$ & & 4.0102 &4.0123 & & & 116.3 &116.1 & \nl
$2s^22p^4(^3P_0)29l\ (l\ge0)$ & & 4.0814 &4.0768 & & & 28.4 &27.5 & \nl
blend & & 4.0086 & 4.0025 & $4.0717\pm0.0006$ & & 162.1 & 168.0 & $159.9\pm 2.8$ \nl
& & & & & & & & \nl
$2s^22p^4(^3P_1)26l\ (l\ge0)$ & & 4.5449 & 4.5455 & $4.5791\pm0.0145$ & &   92.6 & 96.9 &
 $101.0\pm 7.9$ \nl
& & & & & & & &\nl
$2s2p^5(^3P^o_0)6f$ & & 4.8966 &4.9530 & & & 50.0 &50.8 & \nl
$2s^22p^4(^3P_1)27l\ (l\ge0)$ & & 5.0212 &5.0273 & & & 74.7 &80.5 & \nl
blend & & 4.9701 & 4.9986 & $5.0714\pm0.0107$ & &  124.7 & 131.3 &
 $121.2\pm12.9$ \nl
& & & & & & & &\nl
$2s^22p^4(^3P_1)28l\ (l\ge0)$ & & 5.4473 & 5.4508 & $5.4803\pm0.0127$ & &   61.7 & 61.2 &
  $56.3\pm 9.3$ \nl
$2s^22p^4(^1D_2)17l\ (l\ge2)$ & & 5.6385 & 5.6326 & $5.6779\pm0.0035$ & &  296.0 & 299.8 &
 $297.1\pm 8.6$ \nl
$2s^22p^4(^3P_1)29l\ (l\ge0)$ & & 5.8299 & 5.8331 & $5.8797\pm0.0095$ & &   50.9 & 50.6 &
  $83.8\pm 6.1$ \nl
$2s^22p^4(^3P_1)30l\ (l\ge0)$ & & 6.1750 & 6.1780 & $6.2025\pm0.0149$ & &   44.2 & 42.9 &
  $52.9\pm 3.4$ \nl
$2s^22p^4(^3P_1)31l\ (l\ge0)$ & & 6.4987 & 6.4903 & $6.5297\pm0.0187$ & &   37.5 & 37.0 &
  $49.1\pm 4.0$ \nl
$2s^22p^4(^3P_1)32l\ (l\ge0)$ & & 6.7809 & 6.7721 & $6.7801\pm0.0241$ & &   32.6 & 32.5 &
  $41.7\pm 6.2$ \nl
& & & & & & & & \nl
$2s^22p^4(^3P_1)33l\ (l\ge0)$ & & 7.0378 &7.0313 & & & 28.7 &28.3 & \nl
$2s^22p^4(^1D_2)18l\ (l\le1)$ & & 7.0628 &7.0640 & & & 28.2 &31.4 & \nl
blend & & 7.0502 & 7.0485 & $7.0426\pm0.0181$ & & 56.9 & 59.7 & $53.0\pm 6.2$ \nl
& & & & & & & & \nl
$2s^22p^4(^1D_2)18l\ (l\ge2)$ & & 7.2941 & 7.2884 & $7.2885\pm0.0029$ & &  196.3 & 194.2 &
 $143.5\pm10.0$ \nl
$2s^22p^4(^1D_2)19l\ (l\ge1)$ & & 8.6945 & 8.6723 & $8.7245\pm0.0056$ & &  138.0 & 154.3 &
 $142.5\pm 3.4$ \nl
$2s^22p^4(^1S_0)12p$ & & 8.9022 & 8.9037 & $8.9874\pm0.0249$ & & 11.9 & 14.8 &   $14.3\pm  2.7$ \nl
$2s^22p^4(^1S_0)12d$ & & 9.4051 & 9.4051 & $9.4507\pm0.0138$ & & 44.4 & 34.9 &   $27.6\pm  3.1$ \nl
$2s^22p^4(^1S_0)12l\ (l\ge3)$ & & 9.6654 &9.6651 & & & 18.3 &11.6 & \nl
$2s^22p^4(^1D_2)20l\ (l\ge0)$ & & 9.8697 &9.8647 & & & 118.5 &106.2 & \nl
blend & & 9.8424 & 9.8450 & $9.9320\pm0.0066$ & &  136.8 & 117.8 & $97.2\pm 3.8$ \nl
& & & & & & & & \nl
$2s^22p^4(^1D_2)21l\ (l\ge0)$ & & 10.903 & 10.896 & $10.951\pm0.008$  & &   83.4 & 82.8 &
  $66.5\pm 3.3$ \nl
$2s^22p^4(^3P_1)nl\ (34\le n\lesssim 130,\ l\ge0)$ & &  && $7.1-11.7$& 
& 422.2 & 325.0 & $379.6\pm16.5$ \nl
$2s^22p^4(^1D_2)22l\ (l\ge0)$ & & 11.800 & 11.794 & $11.814\pm0.005$  & &   55.4 & 54.0 &
  $51.6\pm 2.1$ \nl
$2s^22p^4(^1D_2)23l\ (l\ge0)$ & & 12.579 & 12.573 & $12.600\pm0.007$  & &   44.9 & 44.2 &
  $37.5\pm 2.1$ \nl
$2s^22p^4(^1D_2)24l\ (l\ge0)$ & & 13.262 & 13.258 & $13.299\pm0.008$  & & 38.0 & 36.5 &
  $34.5\pm 2.1$ \nl
\enddata
\end{deluxetable}

\setcounter{table}{3}
\begin{deluxetable}{ccccccccc}
\footnotesize
\tablecaption{Continued.}
\tablewidth{0pt}
\tablehead{ 
\colhead{ } & \colhead{ } &
\multicolumn{3}{c}{$E_d$ (eV)} & \colhead{ } &
\multicolumn{3}{c}{$\hat{\sigma}_d$ ($10^{-21}$ cm$^2$ eV)} \\
\cline{3-5} \cline{7-9} \\
\colhead{Resonance} & \colhead{ } & 
\colhead{MCDF\tablenotemark{a}} & \colhead{MCBP\tablenotemark{a}} &
\colhead{Experiment\tablenotemark{b,c}}  & 
\colhead{ } & \colhead{MCDF} & \colhead{MCBP} &
\colhead{Experiment\tablenotemark{b}} 
}
\startdata
$2s^22p^4(^1D_2)25l\ (l\ge0)$ & & 13.865&13.862 & & & 31.4 &30.3 & \nl
$2s^22p^4(^1S_0)13d$ & & 13.994 &13.993 & & & 10.1&10.3 & \nl
blend & & 13.897 & 13.895 & $13.903\pm0.009$ & & 41.5 & 40.6 & $36.5\pm 2.2$ \nl
& & & & & & & & \nl
$2s^22p^4(^1S_0)13l\ (l\ge3)$ & & 14.205 &14.203 & & & 7.6 &4.6 & \nl
$2s^22p^4(^1D_2)26l\ (l\ge0)$ & & 14.400 &14.397 & & & 26.9 &25.7 & \nl
blend& & 14.349 & 14.368 & $14.398\pm0.011$ & & 34.5 & 30.3 & $32.0\pm 2.5$ \nl
& & & & & & & & \nl
$2s^22p^4(^1D_2)27l\ (l\ge0)$ & & 14.876 & 14.873 & $14.892\pm0.016$  & &   23.2 & 22.2 &
  $22.8\pm 2.5$ \nl
$2s^22p^4(^1D_2)28l\ (l\ge0)$ & & 15.302 & 15.299 & $15.288\pm0.021$  & &   20.2 & 19.3 &
  $17.5\pm 2.4$ \nl
$2s^22p^4(^1D_2)29l\ (l\ge0)$ & & 15.684 & 15.682 & $15.704\pm0.024$  & &   17.7 & 17.0 &
  $15.9\pm 2.4$ \nl
$2s2p^5(^3P_2^o)7s\ (J=5/2)$ & & 17.890 & 17.943 & $17.912\pm0.066$ & &   22.6 & 21.1 &
  $27.2\pm 2.4$ \nl
$2s2p^5(^3P_2^o)7s\ (J=3/2)$ & & 18.220 & 18.270 & $18.263\pm0.065$ & &   12.7 & 11.9 &
 $15.5\pm  2.4$ \nl
$2s2p^5(^3P_2^o)7p_{1/2}\ (J=3/2,5/2)$& & 20.211 & 20.253 & $20.320\pm0.017$ & & 43.2 & 43.4 &
    $46.1\pm4.4$\nl
$2s2p^5(^3P_2^o)7p_{3/2}\ (J=3/2,5/2,7/2)$& & 20.466 & 20.494 & $20.566\pm0.008$ & & 
106.2 & 114.1 & $110.5\pm7.0$\nl
$2s^22p^4(^1D_2)nl\ (30\le n\lesssim 130,\ l\ge0)$ & & & &$15.8-21.0$
 & & 204.4 & 198.6 & $208.7\pm12.8$ \nl
& & & & & & & & \nl
$2s2p^5(^3P_2^o)7d_{5/2}\ (J=9/2)$ & & 22.785 & & & & 36.5 & & \nl
$2s2p^5(^3P_2^o)7d_{3/2}\ (J=3/2,5/2,7/2)$ & & 22.857 & & & & 81.0 & & \nl
blend & & 22.830 & 22.859 & $22.828\pm0.011$ & & 117.5 & 108.4 &
 $109.0\pm4.1$\nl
& & & & & & & & \nl
$2s2p^5(^3P_2^o)7d_{5/2}\ (J=5/2,7/2)$ & & 23.127& 23.156 &$23.144\pm0.015$ & &54.6 & 51.3 & 
  $63.7\pm3.3$\nl
$2s2p^5(^3P_2^o)7d_{5/2}\ (J=1/2,3/2)$ & & 23.439 & 23.497 & $23.460\pm0.016$ & &53.9& 63.4 & 
  $44.6\pm3.3$\nl
$2s2p^5(^3P^o_2)7f$ & & 24.224& 24.260 &$24.223\pm0.015$ & & 267.8 & 295.4 & $238.1\pm 7.2$ \nl
& & & & & & & & \nl
$2s2p^5(^3P^o_2)7l\ (l\ge4)$ & & 24.410 &24.444 & & & 500.0 &664.0 & \nl
$2s2p^5(^1P_1^o)6s$ & & 24.413 &24.471 & & & 6.8 &6.3 &\nl
blend & & 24.410 & 24.444 & $24.458\pm0.015$ & & 506.8 & 670.3 & $633.1\pm 8.3$ \nl
& & & & & & & & \nl
$2s2p^5(^3P^o_1)7p$ & & 28.082&28.060 & & & 42.9 &45.5 & \nl
$2s2p^5(^1P_1^o)6p$ & & 28.293 &28.293 & & & 12.1 &13.1 & \nl
blend & & 28.142 & 28.112 & $28.082\pm0.012$ & & 55.0 & 58.6 & $47.0\pm 3.4$ \nl
& & & & & & & & \nl
$2s2p^5(^3P^o_1)7d$ & & 30.649 & 30.728 & $30.651\pm0.010$ & &   56.7 & 75.4 &
  $67.5\pm 3.5$ \nl
$2s2p^5(^3P^o_1)7f$ & & 31.899 & 31.925 & $31.901\pm0.027$ & &   65.4 & 101.3 &
  $65.1\pm14.6$ \nl
\enddata
\end{deluxetable}

\setcounter{table}{3}
\begin{deluxetable}{ccccccccc}
\footnotesize
\tablecaption{Continued.}
\tablewidth{0pt}
\tablehead{ 
\colhead{ } & \colhead{ } &
\multicolumn{3}{c}{$E_d$ (eV)} & \colhead{ } &
\multicolumn{3}{c}{$\hat{\sigma}_d$ ($10^{-21}$ cm$^2$ eV)} \\
\cline{3-5} \cline{7-9} \\
\colhead{Resonance} & \colhead{ } & 
\colhead{MCDF\tablenotemark{a}} & \colhead{MCBP\tablenotemark{a}} &
\colhead{Experiment\tablenotemark{b,c}}  & 
\colhead{ } & \colhead{MCDF} & \colhead{MCBP} &
\colhead{Experiment\tablenotemark{b}} 
}
\startdata
$2s2p^5(^3P^o_1)7l\ (l\ge4)$ & & 32.096&32.110 & & & 213.9 &255.0 & \nl
$2s2p^5(^1P_1^o)6d$ & & 32.403 & 32.426& & & 18.3 &19.6 & \nl
blend & & 32.120 & 32.133 & $32.134\pm0.009$ & & 232.2 & 274.6 & $267.2\pm17.6$ \nl
& & & & & & & & \nl
$2s2p^5(^1P^o_1)6l\ (l\ge3)$ & & 34.556 & 34.571 & $34.644\pm0.016$ & &  41.8 & 44.4 &
  $42.6\pm 3.0$ \nl
$2s2p^5(^3P^o_2)8p$ & & 42.833 & 42.873 & $42.936\pm0.010$ & & 45.4 & 47.6 &$42.0\pm 2.2$ \nl
$2s2p^5(^3P^o_2)8d$ & & 44.568 & 44.634 & $44.684\pm0.008$ & & 78.0 & 81.4 & $76.3\pm 2.7$ \nl
$2s2p^5(^3P^o_2)8l\ (l\ge3)$ & & 45.453 & 45.494 & $45.497\pm0.001$ & & 333.2 & 317.1 &
 $329.3\pm 2.4$ \nl
$2s2p^5(^3P^o_1)8p$ & & 50.553 & 50.517 & $50.612\pm0.020$ & & 17.4 & 17.9 & $17.1\pm 2.0$ \nl
$2s2p^5(^3P^o_1)8d$ & & 52.287 & 52.277 & $52.389\pm0.015$ & & 30.8 & 30.5 & $28.1\pm 2.1$ \nl
$2s2p^5(^3P^o_1)8l\ (l\ge3)$ & & 53.170 & 53.165 & $53.310\pm0.003$ & & 140.7 & 137.3 &
 $143.0\pm 2.1$ \nl
$2s2p^5(^3P^o_2)9p$ & & 58.097 & 58.133 & $58.277\pm0.016$ & & 25.7 & 26.7 & $22.3\pm 1.6$ \nl
$2s2p^5(^3P^o_2)9d$ & & 59.276 & 59.359 & $59.485\pm0.011$ & & 48.2 & 46.2 & $35.7\pm 1.7$ \nl
$2s2p^5(^3P^o_2)9l\ (l\ge3)$ & & 59.933 & 59.969 & $60.141\pm0.002$ & & 229.7 & 221.9 &
 $210.0\pm 2.4$ \nl
$2s2p^5(^3P^o_1)9p$ & & 65.793 & 65.786 & $65.938\pm0.031$ & & 10.1 & 10.7 & $12.9\pm 1.8$ \nl
$2s2p^5(^3P^o_1)9d$ & & 66.977 & 67.010 & $67.239\pm0.017$ & & 18.7 & 18.1 & $25.6\pm 1.7$ \nl
$2s2p^5(^3P^o_1)9l\ (l\ge3)$ & & 67.649 & 67.640 & $67.841\pm0.005$ & & 103.5 & 101.3 &
  $95.7\pm 1.9$ \nl
$2s2p^5(^3P^o_2)10p$ & & 68.950 & 68.985 & $69.150\pm0.030$ & & 17.4 & 18.0 &$13.3\pm 1.7$ \nl
$2s2p^5(^3P^o_2)10d$ & & 69.833 & 69.871 &$70.023\pm0.023$ & & 32.5 & 31.2 & $23.0\pm 1.8$ \nl
$2s2p^5(^3P^o_2)10l\ (l\ge3)$ & & 70.286 & 70.321 & $70.534\pm0.003$ & & 182.0 & 177.7 &
 $166.8\pm 2.5$ \nl
$2s2p^5(^3P^o_1)10p$ & & 76.654 &76.643 & & & 7.3 &7.5 & \nl
$2s2p^5(^3P^o_2)11p$ & & 76.941 &76.976 & & & 13.1 &13.4 & \nl
blend & & 76.840 & 76.857 & $76.7-77.5$\tablenotemark{d} & & 20.4 & 20.9 &
 $23.2\pm 2.2$ \nl
& & & & & & \nl
$2s2p^5(^3P^o_2)11l\ (l\ge2)$ & & 77.898 &77.934 & & & 179.1 &176.0 & \nl
$2s2p^5(^3P^o_1)10l\ (l\ge2)$ & & 77.941 &77.936 & & & 97.1 &95.9 & \nl
blend & & 77.913 & 77.935 & $77.5-79.0$\tablenotemark{d} & & 276.2 & 271.9 &
 $230.1\pm 2.7$ \nl
& & & & & & & & \nl
$2s2p^5(^3P^o_2)12p$ & & 82.997 & 83.031 & $83.340\pm0.039$ & & 10.5 & 10.6 &$9.3\pm 1.2$ \nl
$2s2p^5(^3P^o_2)12l\ (l\ge2)$ & & 83.736 & 83.771 & $84.036\pm0.003$ & & 156.8 & 154.9 &
 $135.0\pm 1.8$ \nl
$2s2p^5(^3P^o_1)11p$ & & 84.649 & 84.637 & $84.613\pm0.040$ & & 6.6 & 5.9 & $10.4\pm 1.4$ \nl
$2s2p^5(^3P^o_1)11d$ & & 85.307 & 85.298 & $85.512\pm0.096$ & & 9.7 & 9.8 &  $6.2\pm 2.3$ \nl
$2s2p^5(^3P^o_1)11l\ (l\ge3)$ & & 85.660 & 85.649 & $85.908\pm0.009$ & & 73.0 & 72.7 &  
  $71.4\pm 2.4$ \nl
$2s2p^5(^3P^o_2)13p$ & & 87.629 & 87.727 & $87.832\pm0.063$ & & 10.6 & 8.9 &$10.3\pm 2.3$ \nl
$2s2p^5(^3P^o_2)13l\ (l\ge2)$ & & 88.276 & 88.310 & $88.569\pm0.005$ & & 140.6 & 140.1 &
 $127.9\pm 2.8$ \nl
\enddata
\end{deluxetable}

\setcounter{table}{3}
\begin{deluxetable}{ccccccccc}
\footnotesize
\tablecaption{Continued.}
\tablewidth{0pt}
\tablehead{ 
\colhead{ } & \colhead{ } &
\multicolumn{3}{c}{$E_d$ (eV)} & \colhead{ } &
\multicolumn{3}{c}{$\hat{\sigma}_d$ ($10^{-21}$ cm$^2$ eV)} \\
\cline{3-5} \cline{7-9} \\
\colhead{Resonance} & \colhead{ } & 
\colhead{MCDF\tablenotemark{a}} & \colhead{MCBP\tablenotemark{a}} &
\colhead{Experiment\tablenotemark{b,c}}  & 
\colhead{ } & \colhead{MCDF} & \colhead{MCBP} &
\colhead{Experiment\tablenotemark{b}} 
}
\startdata
$2s2p^5(^3P^o_1)12p$ & & 90.706 & 90.694 & $91.085\pm0.105$ & & 5.4 & 4.6 & $5.5\pm 1.9$ \nl
& & & & & & & & \nl
$2s2p^5(^3P^o_2)14p$ & & 91.410 &91.442 & & & 7.6 &7.7 & \nl
$2s2p^5(^3P^o_1)12l\ (l\ge2)$ & & 91.483 &91.444 & & & 65.6 &73.1 & \nl
blend & & 91.475 & 91.444 & $91.708\pm0.012$ & & 73.2 & 80.8 & $69.2\pm 2.3$ \nl
& & & & & & & & \nl
$2s2p^5(^3P^o_2)14l\ (l\ge2)$ & & 91.877 & 91.911 & $92.184\pm0.007$ & & 129.2 & 128.9 &
$111.3\pm 3.0$ \nl
$2s2p^5(^3P^o_2)15p$ & & 94.402 & 94.436 & $94.636\pm0.095$ & & 6.6 & 6.7 &  $4.0\pm 1.5$ \nl
$2s2p^5(^3P^o_2)15l\ (l\ge2)$ & & 94.782 & 94.815 & $95.083\pm0.005$ & & 118.5 & 118.7 &
 $97.0\pm 1.6$ \nl
$2s2p^5(^3P^o_1)13p$ & & 95.654 & 95.393 & $95.537\pm0.048$ & & 10.4 & 4.0 &$10.2\pm 1.5$ \nl
$2s2p^5(^3P^o_1)13l\ (l\ge2)$ & & 96.015 & 95.982 & $96.268\pm0.007$ & & 59.6 & 66.5 &
 $57.8\pm 1.4$ \nl
$2s2p^5(^3P^o_2)16p$ & & 96.845 & 96.879 & $96.995\pm0.079$ & & 6.0 & 6.0 & $5.5\pm 1.4$ \nl
$2s2p^5(^3P^o_2)16l\ (l\ge2)$ & & 97.158 & 97.192 & $97.485\pm0.005$ & & 111.0 & 111.2 &
 $95.5\pm 1.5$ \nl
$2s2p^5(^3P^o_2)17l\ (l\ge1)$ & & 99.127 & 99.148 & $99.431\pm0.010$ & & 104.2 & 110.4 & 
 $93.3\pm 3.1$ \nl
$2s2p^5(^3P^o_1)14l\ (l\ge1)$ & & 99.610 & 99.560 & $99.874\pm0.015$ & & 64.4 & 64.8 &
 $59.9\pm 3.0$ \nl
$2s2p^5(^3P^o_2)18l\ (l\ge0)$ & & 100.76 & 100.80 & $101.10\pm0.01$  & & 104.4 & 105.8 &
 $91.7\pm 1.9$ \nl
$2s2p^5(^3P^o_2)19l\ (l\ge0)$ & & 102.16 & 102.19 & $102.52\pm0.02$  & & 100.5 & 100.7 & 
 $99.9\pm 6.7$ \nl
$2s2p^5(^3P^o_1)15l\ (l\ge1)$ & & 102.50 & 102.47 & $102.87\pm0.03$  & &  56.7 & 60.4 & 
 $52.2\pm 6.5$ \nl
$2s2p^5(^3P^o_2)20l\ (l\ge0)$ & & 103.35 & 103.39 & $103.69\pm0.01$  & &  91.3 & 95.9 & 
 $76.5\pm 2.5$ \nl
$2s2p^5(^3P^o_2)21l\ (l\ge0)$ & & 104.38 & 104.42 & $104.70\pm0.02$  & &  84.7 & 83.6 & 
 $70.6\pm 5.1$ \nl
$2s2p^5(^3P^o_1)16l\ (l\ge1)$ & & 104.87 & 104.85 & $105.13\pm0.05$  & &  52.9 & 56.4 & 
 $47.0\pm 4.4$ \nl
$2s2p^5(^3P^o_2)22l\ (l\ge0)$ & & 105.27 & 105.31 & $105.60\pm0.02$  & &  80.7 & 79.5 & 
 $74.6\pm 4.2$ \nl
$2s2p^5(^3P^o_2)23l\ (l\ge0)$ & & 106.05 & 106.08 & $106.42\pm0.01$  & &  78.0 & 76.0 & 
 $70.9\pm 2.4$ \nl
& & & & & & & & \nl
$2s2p^5(^3P^o_2)nl\ (24\le n\lesssim 130,\ l\ge0)$ & & 106.7-114.2 & & & 
& 2201.0 &2196.8 & \nl
$2s2p^5(^3P^o_1)nl\ (17\le n\lesssim 130,\ l\ge0)$ & & 106.5-121.9 & & & 
& 995.0 &1031.6 & \nl
blend & & & &$106.5-123.0$ & & 3196.0 & 3228.4 & $2491.9\pm8.3$ \nl
\enddata
\tablenotetext{a}{Weighted energy:
$E_d=\sum_i E_i \hat{\sigma}_i / \sum_i \hat{\sigma}_i$.}
\tablenotetext{b}{1$\sigma$ statistical fitting uncertainties only.}
\tablenotetext{c}{Absolute energy scale uncertainty $\lesssim 0.7\%$.}
\tablenotetext{d}{Unable to fit for resonance energy.}
\end{deluxetable}

\vfill
\clearpage
\eject

\begin{deluxetable}{ccccccccc}
\footnotesize
\tablecaption{Experimental and theoretical quantum defects $\mu_l$ for
Fe XVII and Fe XVIII.  Experimental Fe XVII values are determined using
Fe XVIII to Fe XVII $\Delta n=0$ DR via the $^2P_{3/2}-\ ^2P_{1/2}$ and
$^2P_{3/2}-\ ^2S_{1/2}$ core excitations.  Experimental Fe XVIII values
are determined using Fe XIX to Fe XVIII $\Delta n=0$ DR via the
$^3P_2-\ ^3P_1^o$ and $^3P_2-\ ^3P_2^o$ core excitations.  The
1$\sigma$ statistical fitting uncertainties are given.  Experimental
results are for ions with an excited core configuration.  Theoretical
values are from Theodosiou, Inokuti, \& Manson (1986) and are for ions
with a ground state core configuration.
\label{tab:qd}}
\tablewidth{0pt}
\tablehead{\colhead{ } & \colhead{ } & \multicolumn{3}{c}{Fe XVII} 
& \colhead{ } & \multicolumn{3}{c}{Fe XVIII}\\
\cline{3-5} \cline{7-9}\\
\colhead{$l$} & \colhead{ } & \colhead{$^2P_{3/2}-\ ^2P_{1/2}$} &
\colhead{$^2P_{3/2}-\ ^2S_{1/2}$} & \colhead{Theory} & \colhead{ }
& \colhead{$^3P_2-\ ^3P_1^o$} &
\colhead{$^3P_2-\ ^3P_2^o$} & \colhead{Theory}
}
\startdata
$s$ & & $0.2762\pm0.0006$ &                   & 0.2616 \nl
$p$ & & $0.1786\pm0.0009$ & $0.1811\pm0.0003$ & 0.1718 & & $0.1455\pm0.0025$ 
& $0.1455\pm0.0022$ & 0.1460 \nl
$d$ & & $0.0612\pm0.0013$ & $0.0679\pm0.0002$ & 0.0573 & & $0.0441\pm0.0039$ 
& $0.0478\pm0.0024$ & 0.0502 \nl
\enddata
\end{deluxetable}

\vfill
\clearpage
\eject

\begin{deluxetable}{cccccccccccc}
\footnotesize
\tablecaption{Fit parameters for the experimentally
inferred and theoretical MCDF Fe XVIII to Fe XVII and
Fe XIX to Fe XVIII $\Delta n=0$ DR rate coefficients.
The units for $c_i$ are cm$^3$~s$^{-1}$~K$^{1.5}$ and for
$E_i$ are eV.
\label{tab:fitparameters}}
\tablewidth{0pt}
\tablehead{
\colhead{ } & \multicolumn{5}{c}{Fe XVIII} & \colhead{ } &
\multicolumn{5}{c}{Fe XIX} \\
\colhead{ } & \multicolumn{2}{c}{Experiment} & \colhead{ } &
\multicolumn{2}{c}{MCDF} & \colhead{ } & \multicolumn{2}{c}{Experiment} & 
\colhead{ } & \multicolumn{2}{c}{MCDF}\\
\cline{2-3} \cline{5-6} \cline{8-9} \cline{11-12}\\
\colhead{$i$} & \colhead{$c_i$} & \colhead{$E_i$} &
& \colhead{$c_i$} & \colhead{$E_i$} & \colhead{ } & \colhead{$c_i$} &
\colhead{$E_i$} & & \colhead{$c_i$} & \colhead{$E_i$}}
\startdata
1 & 4.79e-6 & 2.22e-1 & & 5.55e-6 & 2.15e-1 & & 3.73e-5 & 6.61e-2 & 
& 3.16e-5 & 6.19e-2 \nl
2 & 9.05e-5 & 5.24e-1 & & 7.37e-5 & 5.36e-1 & & 1.60e-5 & 2.14e-1 & 
& 2.14e-5 & 1.94e-1 \nl
3 & 3.48e-5 & 1.16e+0 & & 7.47e-5 & 1.95e+0 & & 2.33e-4 & 1.11e+0 & 
& 2.77e-4 & 1.09e+0 \nl
4 & 1.83e-4 & 2.52e+0 & & 2.06e-4 & 3.79e+0 & & 3.63e-4 & 2.60e+0 & 
& 3.97e-4 & 2.50e+0 \nl
5 & 5.26e-4 & 6.57e+0 & & 6.64e-4 & 1.17e+1 & & 1.16e-3 & 6.54e+0 & 
& 1.43e-3 & 6.78e+0 \nl
6 & 2.12e-3 & 1.90e+1 & & 1.27e-3 & 2.18e+1 & & 5.56e-3 & 2.53e+1 & 
& 5.85e-3 & 2.63e+1 \nl
7 & 4.29e-3 & 5.66e+1 & & 5.78e-3 & 6.40e+1 & & 4.12e-2 & 9.70e+1 & 
& 5.18e-2 & 1.00e+2 \nl
8 & 3.16e-2 & 1.21e+2 & & 3.03e-2 & 1.26e+2 & &         &         & 
& & \nl
\enddata
\end{deluxetable}

\vfill
\clearpage
\eject

\clearpage
\eject

\end{document}